\documentclass[hidelinks,onefignum,onetabnum]{siamart220329}

\usepackage{lipsum}
\usepackage{amsfonts}
\usepackage{graphicx}
\usepackage{epstopdf}
\usepackage{algorithmic}
\usepackage{comment}
\usepackage[section]{placeins}
\usepackage{subfig}
\ifpdf
  \DeclareGraphicsExtensions{.eps,.pdf,.png,.jpg}
\else
  \DeclareGraphicsExtensions{.eps}
\fi

\newcommand{\beqsub}{\begin{subequations}}
\newcommand{\eeqsub}{\end{subequations}}

\usepackage{ulem} 

\graphicspath{{./figs/}}

\makeatletter
\providecommand*{\diff}%
{\@ifnextchar^{\DIfF}{\DIfF^{}}}
\def\DIfF^#1{%
	\mathop{\mathrm{\mathstrut d}}%
	\nolimits^{#1}\gobblespace}
\def\gobblespace{%
	\futurelet\diffarg\opspace}
\def\opspace{%
	\let\DiffSpace\!%
	\ifx\diffarg(%
	\let\DiffSpace\relax
	\else
	\ifx\diffarg[%
	\let\DiffSpace\relax
	\else
	\ifx\diffarg\{%
	\let\DiffSpace\relax
	\fi\fi\fi\DiffSpace}

\providecommand*{\deriv}[3][]{%
	\frac{\diff^{#1}#2}{\diff #3^{#1}}}

 \renewcommand{\exp}[1]{\textrm{e}^{{#1}}}

\newcommand{\Sfd}{S_{\rm{fd}}}
 \newcommand{\Ro}{\mathcal{R}_0}
 \newcommand{\Rf}{\mathcal{R}_{\rm f}}
 \newcommand{\Rfhat}{\hat{\mathcal{R}}_{\rm f}}
 \newcommand{\Rp}{\mathcal{R}_{\rm p}}

\newsiamremark{remark}{Remark}
\newsiamremark{hypothesis}{Hypothesis}
\crefname{hypothesis}{Hypothesis}{Hypotheses}
\newsiamthm{claim}{Claim}

\headers{Behaviour response to novel and established diseases}{A. Kaur, R. C. Tyson, and I. R. Moyles}

\title{The impact of fear and behaviour response to established and novel diseases \thanks{Submitted to the editors DATE.
\funding{This work was funded by the Natural Sciences and Engineering Research Council of Canada through the Discovery Grant and Emerging and Infectious Disease Modelling programs.}}}

\author{Avneet Kaur\thanks{Department of Computer Science, Mathematics, Physics, and Statistics, University of British Columbia, Kelowna, BC 
  (\email{thisisavneet0001@gmail.com}, \email{rebecca.tyson@ubc.ca}).}
\and Rebecca C. Tyson \footnotemark[2]
\and Iain R. Moyles\thanks{Department of Mathematics and Statistics, York University, Toronto, ON (\email{imoyles@yorku.ca}, \textit{Corresponding Author}).}}

\usepackage{amsopn}

\ifpdf
\hypersetup{
  pdftitle={An Example Article},
  pdfauthor={D. Doe, P. T. Frank, and J. E. Smith}
}
\fi

\begin{document}

\maketitle

\begin{abstract}
We analyze a disease transmission model that allows individuals to acquire fear and change their behaviour to reduce transmission. Fear is acquired through contact with infected individuals and through the influence of fearful individuals. We analyze the model in two limits: First, an Established Disease Limit (EDL), where the spread of the disease is much faster than the spread of fear, and second, a Novel Disease Limit (NDL), where the spread of the disease is comparable to that of fear. For the EDL, we show that the relative rate of fear acquisition to disease transmission controls the size of the fearful population at the end of a disease outbreak, and that the fear-induced contact reduction behaviour has very little impact on disease burden. Conversely, we show that in the NDL, disease burden can be controlled by fear-induced behaviour depending on the rate of fear loss. Specifically, fear-induced behaviour introduces a contact parameter $p$, which if too large prevents the contact reduction from effectively managing the epidemic. We analytically identify a critical prophylactic behaviour parameter $p=p_c$ where this happens leading to a discontinuity in epidemic prevalence. We show that this change in disease burden introduces delayed epidemic waves.
\end{abstract}



\section{Introduction}

Human behavioural changes play a crucial role in shaping the trajectory of infectious diseases. As the disease pathogen spreads, the associated morbidity and mortality give rise to population-level perceptions of the risk and severity of the disease. As a result, people try to minimise those risks by spontaneously changing their behaviours, even in the absence of government directives~\cite{baumgaertner2018influence,baumgaertner2020risk}.

Most commonly, risk minimisation is done through prophylactic actions such as social distancing, wearing face masks, etc. Modelling behaviour and opinion dynamics alone can lead to rich patterns in behaviour adoption and complex dynamics such as the degree to which opinion strength affects behaviour~\cite{Baumgaertner2018,Baumgaertner2016}. However, when coupled with disease dynamics, behaviour can shape the course of the epidemic.  This observation has led to the important field of Behavioural Epidemiology which emphasizes the incorporation of theories from sociology and psychology as key components disease models~\cite{bauch2013behavioral, glaubitz2020oscillatory}. Several reviews including behaviour and its importance in infectious disease modelling have been written~\cite{Funk2010,weston2018infection}.

Many modelling efforts in Behavioural Epidemiology have demonstrated that spontaneous behavioural responses to increased disease prevalence can result in disease dynamics that differ substantially from those predicted by models that do not include human behaviour. For example, incorporating prophylactic behaviour in the response to an emerging disease can induce multiple waves of infection~\cite{moyles2021cost,Tyson2022,iwasa2023waves}. Furthermore, including behaviour in disease models has produced mechanisms that could explain trends observed in data. For example, a model of individual health decisions captured face mask usage during the 2003 SARS epidemic in Hong Kong~\cite{durham2012incorporating}, a model of risk perception was able to explain a sudden change in disease incidence for the 2009 H1N1 epidemic in Italy~\cite{poletti2011effect}, and a model of perceived immunity was able to explain the fall 2021 wave of COVID-19 in Ontario Canada~\cite{molla2023}. It has even been demonstrated that incorporating the planning horizon time into behavioural decision making can itself impact disease burden~\cite{nardin2016planning}.

Naturally, as the inclusion of behaviour strengthened epidemiological predictions, the structure of these mathematical models and the mechanisms that affect disease burden became an area of interest. For example, an adaptation of the classic Susceptible-Infected-Recovered (SIR) formulation of Kermack and McKendrick~\cite{kermack1927contribution} introduced fear as a second contagion and showed that this dual-contagion model can lead to multiple epidemic waves \cite{epstein2008coupled}. Perra et al (2011) generalized this model framework and showed that when fear spreads at a rate faster than disease, the fear-induced contact reduction reduces the final epidemic size (i.e., the total infected population)~\cite{perra2011towards} . A similar bifurcation in final epidemic size was observed in Poletti et al (2011)~\cite{poletti2011effect} where they showed that behaviour changes driven by a high perception of risk of disease, fast adoption of fear, and a relatively slow fear loss, can significantly change the disease dynamics, particularly sudden changes from slow to fast growth in incidence case numbers. 

The complexity of the behaviourally-mediated disease dynamics has limited model analysis. Perra et al (2011) generalized some fear-responsive models and used linear theory to identify some bifurcations in final sizes of fear compartments~\cite{perra2011towards}. However, their most interesting observation was a discontinuity in epidemic prevalence when a self-reinforcement mechanism of fear loss was included in their model. The authors concluded that the mechanism of this bifurcation was outside of linear theory and left it unresolved. A similar fear-loss mechanism was introduced by Epstein et al where they observed bifurcations in epidemic prevalence, but through simulations only \cite{epstein2021triple}. Yet another bifurcation in final epidemic size was observed in the qualitative analysis of a slightly different model~\cite{perra2013modeling}. This model included imitative fear loss, which corresponds to a fearful population losing fear upon interaction with regular susceptibles ($S$), who are unafraid. Overall, the relationship between different disease dynamics, particularly the bifurcations that separate them, and the underlying parameters that drive them is key ingredient in developing our understanding of epidemics and for improving public health interventions.


We consider a fear-incorporating disease model under two disease scenarios, one where the circulating disease is established (e.g., the seasonal flu) and one where the circulating disease is novel (e.g., COVID-19 in early 2020). Using asymptotic and weakly non-linear analysis we determine the role of behaviour in shaping disease burden outcomes for each scenario, including the resolution of discontinuities in epidemic prevalence first observed in \cite{perra2011towards}.

The outline of our article is as follows. In \Cref{sec:model} we define and non-dimensionalize the fear-disease model, determine the established and novel disease limits, and discuss equilibria and stability. We analyse the Established Disease Limit in \Cref{sec:established} and the Novel Disease Limit in \Cref{sec:novel}. We discuss the implications of our results and offer conclusions in \Cref{sec:Discussion}.

\section{Model description}\label{sec:model}

\begin{figure}
    \centering
    \includegraphics[width=0.5\textwidth]{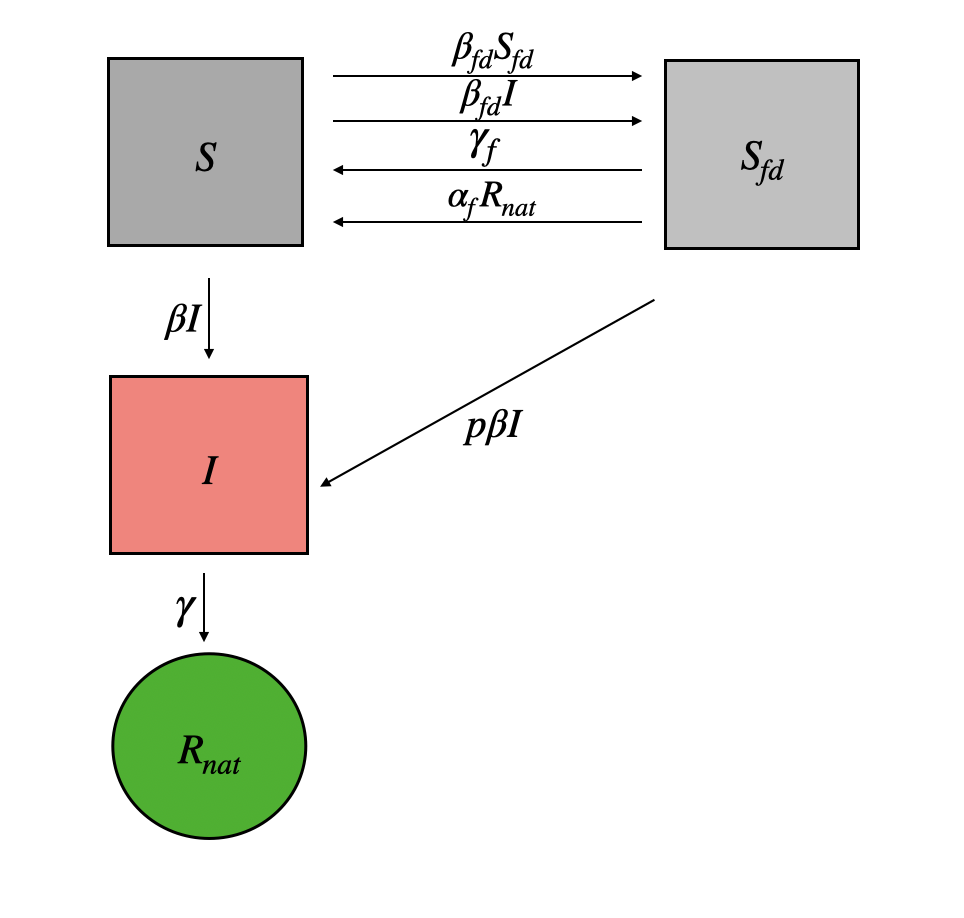}
    \caption{Compartmental diagram of the model~\cite{epstein2021triple}. The compartments $S$ and $S_{\rm{fd}}$ refer to the fearless and disease-fearful susceptibles, respectively. The compartment $I$ represents the infectious individuals, and the compartment $R_{\rm{nat}}$ refers to the individuals who recovered naturally from the disease. The various transition rates between compartments appear above each arrow.}
    \label{fig:model}
\end{figure}

The model we consider has developed from a series of papers (cf. \cite{epstein2008coupled,perra2011towards,epstein2021triple}); we consider the version from Epstein et al (2011)~\cite{epstein2021triple}. The model is an extension of the fundamental Susceptible-Infected-Recovered model~\cite{kermack1927contribution} in which the susceptible compartment is split into two groups: the susceptibles who are fearless ($S$) and fearful ($S_{\rm{fd}}$) of the disease. The compartment $I$ refers to the infectious individuals, and $R_{\rm{nat}}$ refers to the individuals who recovered naturally from the disease. The compartmental diagram of the model is shown in \cref{fig:model}.



We assume that each state variable, $X \in [S, S_{\rm{fd}}, I, R_{\rm{nat}}]$, in our model represents the corresponding fraction of the total population, hence $X\in[0,1]$ and are subject to the constant population constraint $S + S_{\rm{fd}} + I + R_{\rm{nat}} = 1$. Assuming a well-mixed population, the differential equations corresponding to \cref{fig:model} are
\beqsub
\label{sys:full_model}
\begin{align}
    \deriv{S}{t} &= -\beta I S - \beta_{\rm{fd}}(S_{\rm{fd}}+I)S + \gamma_{\rm{f}} S_{\rm{fd}} + \alpha_{\rm{f}} (1 - S - S_{\rm{fd}} - I) S_{\rm{fd}}, \label{eqn:fullS}\\
    \deriv{S_{\rm{fd}}}{t} &= -p\beta I S_{\rm{fd}} + \beta_{\rm{fd}}(S_{\rm{fd}}+I)S - \gamma_{\rm{f}} S_{\rm{fd}} -\alpha_{\rm{f}} (1 - S - S_{\rm{fd}} - I) S_{\rm{fd}},\label{eqn:fullSfd}\\
    \deriv{I}{t} &= \beta I S + p\beta I S_{\rm{fd}} - \gamma I, \label{eqn:fullI}\\
    \deriv{R_{\rm{nat}}}{t} &= \gamma I.\label{eqn:fullR}
\end{align}
\eeqsub
where $\beta$ is the disease transmission rate, $\beta_{\rm fd}$ the fear transmission rate, $\gamma$ the disease recovery rate, $\gamma_{\rm f}$ the rate of spontaneous loss of fear, $\alpha_{\rm f}$ the rate of loss of fear through contact, and $p$ the relative reduction in rate of disease transmission for fearful susceptibles.

The role of fear impacts the model \cref{sys:full_model} in four ways. There are two mechanisms of fear acquisition which come from susceptible individuals interacting with either infected individuals or those susceptible but fearful. There are also two mechanisms of fear loss with a natural loss through which fearful susceptibles lose their fear and a self-reinforcing loss from fearful susceptible individuals who interact with those who have recovered from the disease. We consider this second mechanism to be fear loss due to complacency, as the observation of those who have recovered from the disease leads to a lower perception of risk and reduced prophylactic behaviour. It is tuning of this mechanism that yielded discontinuous epidemic prevalence in other studies \cite{perra2011towards,epstein2021triple}. Numerical study of the model \cref{sys:full_model}
demonstrated that the final epidemic size  increases suddenly as either $p$ or $\alpha_{\rm{f}}$ increase past a critical value~\cite{epstein2021triple}. 
This same study also showed that, as $p$ increases, i.e., as fear-induced contact reduction becomes less effective, the infection dynamics transition from a pair of waves to a single small infection wave.

\subsection{Steady State Analysis}\label{sec:steady}

We review some of the rich aspects of the phase space of \cref{sys:full_model} as first noted by Perra et al (2011)~\cite{perra2011towards}. Particularly, we note that \cref{sys:full_model} permits two classes of equilibria,
\beqsub
\begin{align}
    E_0(S^*) =& (S^*, 0, 0, 1-S^*),\label{eqn:equil1}\\
    E_1(\Sfd^*) =& (\bar{S}, \Sfd^*, 0, 1-\bar{S}-\Sfd^*), \label{eqn:equil2}
\end{align}
\eeqsub
where, 
\begin{equation}\label{eqn:S_bar}
    \bar{S} = \frac{\gamma_{\rm{f}} + \alpha_{\rm{f}}(1-S_{\rm{fd}}^*)}{\alpha_{\rm{f}} + \beta_{\rm{fd}}},
\end{equation} 
and $S^*$ and $S_{\rm{fd}}^*$ are the fractions of the population in the regular susceptible group ($S$), and the fearful susceptible group ($S_{\rm{fd}}$) at equilibrium. As is often the case in disease models, these values cannot be determined uniquely via steady-state analysis as they are dependent on the initial conditions of the model.

The equilibrium $E_0$ \cref{eqn:equil1} has no fearful susceptible individuals, no infected individuals, and a recovered population of $1-S_0^*$. When $S_0^*=1$ then there is no recovered population and the equilibrium $E_0(1)$ is classically known as a \textit{disease-free equilibrium} (DFE)~\cite{van2017reproduction} since the disease never spread.  When $S_0^*\neq0$ then this equilibrium represents a population that has recovered from the disease and retains no fear thereof. We therefore call equilibrium \cref{eqn:equil1} the \textit{fear-free equilibrium}.  

The second equilibrium, $E_1$ \eqref{eqn:equil2}, differs from $E_0$ in that it has a non-zero fearful susceptible population. We therefore call this equilibrium the \textit{fear-endemic equilibrium}. We note that it is possible to have a fear-endemic disease-free equilibrium meaning that fear spreads even when no disease is present. This occurs if $\Sfd^*+\bar{S}=1$ as there are no recovered individuals in the system indicating that the disease did not spread.

\subsubsection{Stability of the Disease-Free Equilibrium} To investigate the stability of the DFE $E_0(1) = (1, 0, 0, 0)$, we obtain the eigenvalues of the Jacobian around the DFE. The Jacobian has a zero eigenvalue and two non-zero eigenvalues. The zero eigenvalue is related to the degeneracy $E_0(S^*)$ for multiple values of $S^*$ and does not affect stability of the DFE to large perturbations. Hence, we consider the non-zero eigenvalues of the Jacobian 
\beqsub
\begin{align}
    \lambda_1 &= \beta - \gamma,\label{eqn:E0(1)_lambda_1}\\
    \lambda_2 &= \beta_{\rm{fd}} -\gamma_{\rm{f}}. \label{eqn:E0(1)_lambda_2}
\end{align}
\label{eqn:lambdaDFE}
\eeqsub
From here we can define a disease basic reproduction number $\Ro$ and a fear basic reproduction number $\Rf$,
\begin{align}
    \Ro=\frac{\beta}{\gamma},\quad
    \Rf=\frac{\beta_{\rm fd}}{\gamma_{\rm f}}, \label{sys:repnums}
\end{align}
such that the DFE is unstable if either $\Ro>1$ or $\Rf>1$. These reproduction numbers could also be derived using the next generation method~\cite{van2002reproduction}.

We are interested in understanding the dynamics of the system when both the disease and fear contagions can spread in the population, so we assume that both reproduction numbers in  \cref{sys:repnums} are greater than one so that the eigenvalues \cref{eqn:lambdaDFE} are positive.

\subsubsection{Stability of the Fear-Free Equilibrium} To determine the stability of the fear-free equilibrium when the disease has spread, i.e., $E_0(S^*)$, $S^*<1$ in \cref{eqn:equil1}, we obtain the non-zero eigenvalues of the Jacobian at $E_0(S^*)$: 
\beqsub
\begin{align}
    \lambda_1 &= (S^* \Ro - 1)\gamma \label{eqn:E0(S0)_lambda_1},\\
    \lambda_2 &= (S^* \Rf - 1)\gamma_{\rm{f}} - (1-S^*)\alpha_{\rm{f}}. \label{E0(S0)_lambda_2}
\end{align}
\eeqsub
Both of these eigenvalues are negative if 
\begin{align}
    S^* < \min\left(\frac{1}{\Ro},\frac{a + 1}{a + \Rf}\right);\qquad  a = \frac{\alpha_{\rm{f}}}{\gamma_{\rm{f}}}.\label{eqn:Sstarmin}
\end{align}
In the absence of fear loss due to complacency (when $a = 0$), condition \cref{eqn:Sstarmin} reduces to
\[
\max(\Ro^{\rm eff},\Rf^{\rm eff})<1,
\]
where $\mathcal{R}_i^{\rm eff}=S^*\mathcal{R}_i$ are the effective reproduction numbers of the final fearless susceptible population $S^*$ and appear regularly in disease models as a condition of reaching herd immunity~\cite{garnett2002,garnett2005}.

\subsubsection{Stability of the Fear-Endemic Equilibrium} The non-zero eigenvalues of the Jacobian at the fear-endemic equilibrium, $E_1(S_{\rm{fd}}^*)$ from \cref{eqn:equil2}, are:
\beqsub
\begin{align}
    \lambda_1 &= - \Rf \gamma_{\rm{f}} S_{\rm{fd}}^* \label{eqn:E1(S_fd)_lamda_1},\\
    \lambda_2 &= \left[\Ro S_{\rm{fd}}^* \left(p - \frac{a}{a + \Rf}\right) + \Ro\left(\frac{a + 1}{a + \Rf}\right) - 1\right] \gamma. \label{eqn:E1(S_fd)_lamda_2}
\end{align}
\eeqsub
The first of these eigenvalues is always negative, but there are a myriad of stability conditions for \cref{eqn:E1(S_fd)_lamda_2}. In the absence of fear loss due to complacency ($a=0$) then \cref{eqn:E1(S_fd)_lamda_2} becomes
\begin{align}
\lambda_2=\left[\Ro\Sfd^*p+\frac{\Ro}{\Rf}-1\right]\gamma.\label{eqn:E1lam2simp}
\end{align}
The first term, $\Ro\Sfd^*p$ is another effective reproduction number where $p$, the modified contact parameter, reduces the disease transmissibilty for susceptible people who are fearful. From \cref{eqn:E1lam2simp} we immediately see that if $\Ro>\Rf$ then the fear-endemic equilibrium $E_1(\Sfd^*)$ is unstable. This result is intuitive because, in this case, the growth rate of the disease is larger than that of fear. However, even if $\Rf>\Ro$, the fear-endemic equilibrium can still be unstable if the behaviour-modified effective reproduction number $\Ro\Sfd^*p$ is large enough, which happens if the fearful group has only a weak prophylactic response (large $p$) and/or a large population size. Thus, even if fear grows quickly, if the prophylactic behaviour that is generated from that fear is not strong enough, the equilibrium can become unstable.

Overall, the stability criteria for the steady states clearly demonstrates that rich dynamics are possible and the non-linear responses observed by Perra et al (2011) are unsurprising \cite{perra2011towards}. This motivates the deeper non-linear analysis of these models which we now perform.
\subsection{Model Scaling}
\label{sec:model_scaling}

The steady state analysis above uncovered key parameter groupings, particularly the reproduction numbers $\mathcal{R}_i$~\cref{sys:repnums} and the fear-loss ratio, $a$~\cref{eqn:Sstarmin}. We are therefore motivated to non-dimensionalize the model \cref{sys:full_model}. We scale time $t\sim\gamma^{-1}$ with the disease recovery time to produce the non-dimensional model
\begin{subequations}\label{sys:ndmod}
    \begin{align}
        \dot{S}=&-\Ro SI-\delta\left[\Rf(\Sfd+I)S-\Sfd-a(1-S-\Sfd-I)\Sfd\right],\\
        \dot{S}_{\rm fd}=&-\Ro p\Sfd I+\delta\left[\Rf(\Sfd+I)S-\Sfd-a(1-S-\Sfd-I)\Sfd\right],\\
        \dot{I}=&\Ro\left(S+p\Sfd\right)I-I,\label{eqn:ndmodI}
    \end{align}
\end{subequations}
where the over-dot indicates differentiation with respect to non-dimensional time. The reproduction numbers and fear-loss ratio appear in this model along with a new non-dimensional parameter
\begin{align}
    \delta=\frac{\gamma_{\rm f}}{\gamma},\label{eqn:deltadef}
\end{align}
which is the ratio of the rates of recovery for fear and disease. We make the assumption that $\delta\ll1$ meaning that the average natural recovery time for fear is much longer than that of the disease. This assumption is supported through literature where studies on fear extinction have indicated that therapy is often required to eliminate fear and that there can be high rates of relapse~\cite{carpenter2019,raeder2020,myers2006different}. Conversely, a study of elderly respondents with a common cold showed that only 15\% had not recovered within 8 days \cite{kim2021}. Similarly, a study on COVID-19 found that 80\% of people had recovered within one month \cite{liu2021}. The recovery rate $\gamma$ in our model represents the timescale over which someone is no longer infectious which is likely faster than the time it takes for a patient to self-declare as being recovered.  Recovery studies thus likely overestimate the loss of infectiousness time (recovery time in our context). 

For the non-dimensional model we take as our initial condition a susceptible population that is entirely fearless toward the disease, and a small group of infected individuals. All other subgroups have zero population.  Thus, we take
\begin{align}
    S(0)=1-\delta I_0,\quad \Sfd(0)=0,\quad I(0)=\delta I_0,\label{eqn:ICstart}
\end{align}
noting that there is no loss of generality if we allow the initial population to include some disease-fearful individuals, i.e. that $\Sfd(0)>0$. Based on the initial conditions \cref{eqn:ICstart}, it is natural to transform the variables as
\begin{align}
    S=1+\delta x,\quad \Sfd=\delta y,\quad I=\delta z,\label{eqn:xyz_intro}
\end{align}
leading to the new problem
\begin{subequations}\label{sys:xyz_base}
    \begin{align}
        \dot{x}=&-\Ro z(1+\delta x)-\delta\left[\Rf(y+z)(1+\delta x)-y-a(x+y+z)y\right],\\        
        \dot{y}=&-\delta \Rf p yz + \delta\left[\Rf(y+z)(1+\delta x)-y-a(x+y+z)y\right],\\
        \dot{z}=&\Ro(1+\delta x+\delta p y)z-z.
    \end{align}
\end{subequations}
When all non-dimensional parameters are $\mathcal{O}(1)$ then since $\delta\ll 1$ the leading order problem is
\begin{subequations}
    \begin{align}
        \dot{x}=&-\Ro z,\\
        \dot{y}=&0,\\
        \dot{z}=&(\Ro-1)z,
    \end{align}
\end{subequations}
and so we see that there is minimal growth in the fearful susceptible population and that susceptibles quickly become infected,
\begin{align}
z(t)=I_0\exp{(\Ro-1)t}\label{eqn:expzbase}
\end{align}
with an outbreak occurring under the assumption that $\Ro>1$. In this limit the fear-reproduction number is comparable to the disease reproduction number meaning that fear acquisition does not occur exceptionally quickly. As such we refer to this as the \textit{Established Disease Limit} (EDL).

Fear will grow slowly in \cref{sys:xyz_base} unless $\Rf\gg1$ indicating that there is a rapid acquisition of fear. In this case we can write, $\Rf=\delta^{-1}\Rfhat$ with $\Rfhat=\beta_{\rm fd}\gamma^{-1}$. Similarly scaling, $a=\delta^{-1}\hat{a}$ with $\hat{a}=\alpha_{\rm f}\gamma^{-1}$ changes the leading order of \cref{sys:xyz_base} to
\begin{subequations}
    \begin{align}
        \dot{x}=&-\Ro z-\Rfhat(y+z),\\
        \dot{y}=&\Rfhat(y+z),\\
        \dot{z}=&(\Ro-1)z.
    \end{align}
\end{subequations}
Once again the disease grows exponentially via \cref{eqn:expzbase}, but now the fearful susceptible population grows with solution
\begin{align}
y(t)=\frac{I_0\Rfhat}{1-\Ro+\Rfhat}\left(\exp{\Rfhat t}-\exp{(\Ro-1)t}\right),\label{eqn:novel_y_growth}
\end{align}
and people become rapidly afraid. In fact, if $\Rfhat>\Ro-1$ then they become afraid more rapidly than the disease grows. We call the scenario where $\Rf\sim\mathcal{O}(\delta^{-1})$ the \textit{Novel Disease Limit} (NDL) since fear acquisition is very rapid suggesting that there is limited population exposure to the pathogen. An example of such a disease would be COVID-19 in early 2020 before it approached endemicity. We analyze these two limits separately and investigate the impact of fear and behaviour control on disease outcomes.

\section{The Established Disease Limit (EDL)}
\label{sec:established}

As was outlined in \Cref{sec:model_scaling} the EDL is one in which each reproduction number leads to an instability in the disease and fear-free equilibrium $\mathcal{R}_i>1$, but that $\mathcal{R}_i\sim\mathcal{O}(1)$. In this case the fear grows slowly while the disease grows rapidly and therefore to analyze the model we consider the initial conditions $S(0)=1-I_0$, $\Sfd(0)=0$, and $I(0)=I_0$ noting that allowing an initial fearful susceptible population does not alter the fundamental structure of the results that follow. We take $S=x$, $\Sfd=\delta y$, and $I=z$ and write \cref{sys:ndmod} as
\begin{subequations}\label{sys:etablishedmod}
    \begin{align}
        \dot{x}=&-\Ro xz-\delta\left[\Rf(\delta y+z)x-\delta y-\delta a(1-x-\delta y-z)y\right],\\
        \dot{y}=&-\Ro p yz + \Rf(\delta y+z)x-\delta y-\delta a(1-x-\delta y-z)y,\\
        \dot{z}=&\Ro(x+\delta py)z-z.
    \end{align}
\end{subequations}
We form an asymptotic expansion,
\begin{align}
    x\sim x_0+\delta x_1+\dots,\quad y\sim y_0+\delta y_1+\dots,\quad z\sim z_0+\delta z_1+\dots,
\end{align}
and to leading order \cref{sys:etablishedmod} becomes
\begin{subequations}
    \begin{align}
        \dot{x}_0=&-\Ro x_0z_0\label{eqn:x0de}\\
        \dot{y}_0=&\Rf x_0z_0-\Ro p y_0z_0,\label{eqn:y0de}\\
        \dot{z}_0=&(\Ro x_0-1)z_0,\label{eqn:z0de}
    \end{align}
\end{subequations}
where we see that the fearful susceptible population has decoupled from the susceptible and infected group. Because of this, the leading order disease dynamics follow a traditional SIR model which has an analytic solution (see for example \cite{brauer2008}),
\begin{align}
    z(x)=1-x_0+\frac{1}{\Ro}\log\left(\frac{x_0}{1-I_0}\right),\label{eqn:SIRsol}
\end{align}
accompanied by the classic SIR results such as single-peak epidemic dynamics. The leading order fearful susceptible population is determined by solving \cref{eqn:y0de} yielding,
\begin{align}
    y_0=\frac{\Rf}{\Ro(1-p)}x_0^p\left((1-I_0)^{1-p}-x_0^{1-p}\right).\label{eqn:y0sol}
\end{align}
We note that \cref{eqn:SIRsol} only provides an implicit relationship between $x_0$ and $z_0$, but this can be used to furnish integral solutions explicit on time. However, since the dynamics mimic the SIR model, the main outstanding issue is the final size of the fearful susceptible group since we showed in \Cref{sec:model} that the equilibria permit both fearless and fearful possibilities. As such, we are primarily concerned with the final size of each compartment. There is no endemic state from the disease due to an absence of waning immunity and so from both \cref{eqn:SIRsol} and the equilibrium discussion in \Cref{sec:model} then $z_0^\infty=\lim_{t\to\infty}z_0=0$. We can add \cref{eqn:x0de} and \cref{eqn:z0de} to determine that the final size for the susceptible group. $x_0^\infty=\lim_{t\to\infty}x_0$, satisfies,
\begin{align}
    \frac{1}{\Ro}\log\left(\frac{1-I_0}{x_0^\infty}\right)=1-x_0^\infty,\label{eqn:x0inf}
\end{align}
and then we can use this value in \cref{eqn:y0sol} to determine $y_0^\infty$. We thus far have that at steady state $S\sim x_0^\infty$ while $\Sfd\sim \delta y_0^\infty$ and we therefore require $x_1^\infty$ so that both final sizes are known to the same order. The $\mathcal{O}(\delta)$ problem from \cref{sys:etablishedmod} is
\begin{subequations}\label{sys:establishedOdel}
    \begin{align}
        \dot{x}_1=&-\Ro(x_1z_0+x_0z_1)-\Rf x_0z_0,\label{eqn:x1de}\\
        \dot{y}_1=&-\Ro p(y_1z_0+y_0z_1)+\Rf (x_1z_0+x_0z_1+x_0y_0)-y_0-a(1-x_0-z_0)y_0,\label{eqn:y1de}\\
        \dot{z}_1=&(\Ro x_0-1)z_1+\Ro x_1z_0+\Ro p y_0z_0,\label{eqn:z1de}
    \end{align}
\end{subequations}
subject to $x_1(0)=y_1(0)=z_1(0)=0$. Adding \cref{eqn:x1de,eqn:z1de} and integrating to infinite time yields,
\begin{align}
    x_1^\infty+y_0^\infty=-\int_0^\infty z_1\diff t.
\end{align}
If we write $x_1=x_0u$ then we can transform \cref{eqn:x1de} to
\[
\dot{u}=-\Ro z_1-\Rf z_0,\qquad u(0)=0,
\]
which upon integrating over infinite time yields
\[
u^\infty=\Ro(x_1^\infty+y_0^\infty)+\Rf(x_0^\infty-1),
\]
where we have used that
\[
\int_0^\infty z_0=1-x_0^\infty.
\]
Finally then we have that
\begin{align}
    x_1^\infty=\left(\frac{\Ro y_0^\infty-\Rf(1-x_0^\infty)}{1-\Ro x_0^\infty}\right)x_0^\infty.
\end{align}
\subsection{Long-time limit}\label{sec:established_long}
The final-size prediction for the fearful susceptible population is $\Sfd\sim\delta y_0$  with $y_0$ given by \cref{eqn:y0sol}. This is positive which implies that there are always fearful susceptibles, something we know not to be the case from the equilibrium considerations in \Cref{sec:model}. It turns out (see \cref{app:established_break}) that $\Sfd$ grows without bound and thus eventually violates the assumption that $\Sfd\sim\mathcal{O}(\delta)$. This happens in a timescale $t\sim\mathcal{O}(\delta^{-1})$ and so it suggests that we let $t=\delta^{-1}\tau$ and write \cref{sys:full_model} in this new timescale as,
\begin{subequations} \label{sys:established_tau}
    \begin{align}
        S_\tau=&\Sfd+a(1-S-\Sfd)\Sfd-\Rf\Sfd S,\\
        {\Sfd}_\tau=&\Rf\Sfd S-\Sfd-a(1-S-\Sfd)\Sfd,\label{eqn:Sfdtaude}
    \end{align}
\end{subequations}
with initial conditions $S(0)=x_0^\infty+\delta x_1^\infty$ and $\Sfd(0)=\delta y_0^\infty$. In writing \cref{sys:established_tau} we have used extinction of the disease to eliminate $I$ in the long-time limit (see \cref{app:established_break}). Thus, fear remains the only contagious agent.

\begin{figure}
    \centering
    \subfloat[Vary $a$, fix $\Rf$ ($\Rf=8$). \label{fig:fearbifa}]
    {
    \includegraphics[width=0.45\textwidth]{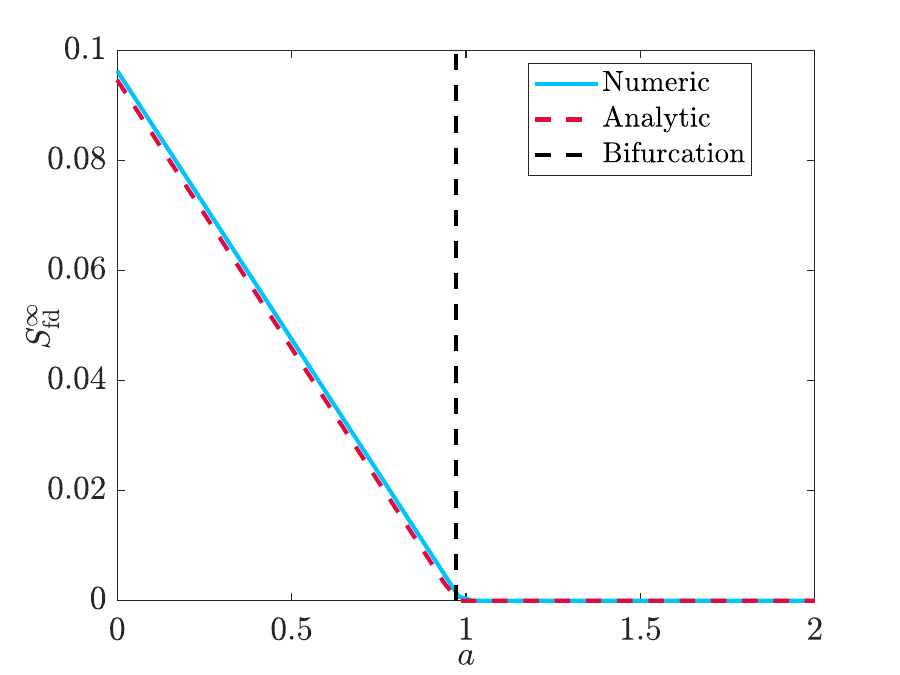}
    }
    \qquad
    \subfloat[Vary $\Rf$, fix $a$ ($a=0$). \label{fig:fearbifRf}]
    {
    \includegraphics[width=0.45\textwidth]{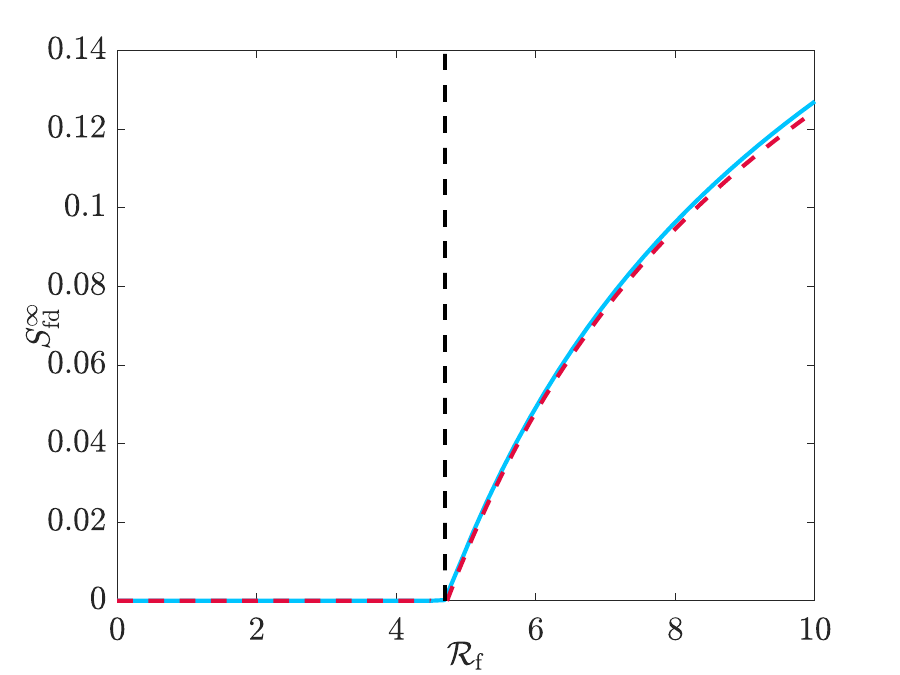}
    }
    \caption{Computation of $\Sfd^\infty=\lim_{t\to\infty}\Sfd(t)$ via simulation of \cref{sys:ndmod} with parameters $\Ro=2$, $p=0.25$, and $\delta=0.01$. The initial conditions are $S=0.99$, $\Sfd=0$, and $I=0.01$. A bifurcation occurs when $\kappa=1$ as defined by \cref{eqn:Sfdtau}. When $\kappa<1$ the predicted final size of the fearful susceptible group is $\Sfd^\infty=0$ while for $\kappa>1$ it is given by \cref{eqn:Sfdinf}. The results from the simulation are plotted as a blue solid line, the predictions from the asymptotic methods are plotted as a red dashed line and the predicted bifurcation point is a vertical black dashed line. }
    \label{fig:fearbif}
\end{figure}

We add the component equations in \cref{sys:established_tau}, integrate, and use the initial conditions to conclude that
\begin{align}
    S+\Sfd=\eta;\qquad \eta=x_0^\infty+\delta(x_1^\infty+y_0^\infty).
\end{align}
From this we can eliminate $S$ in \cref{eqn:Sfdtaude} to determine the long-time fearful susceptible population,
\begin{align}
    \Sfd(\tau)=\frac{(\kappa-1)\delta y_0}{\delta\Rf y_0-(\Rf\delta y_0+1-\kappa)\exp{(1-\kappa)\tau}};\qquad \kappa=\Rf\eta-a(1-\eta),\label{eqn:Sfdtau}
\end{align}
If $\kappa<1$ then the exponential in the denominator of \cref{eqn:Sfdtau} grows as $\tau\to\infty$ so $\Sfd\to0$. Conversely, if $\kappa>1$ then the exponential decays and
\begin{align}
    \Sfd^\infty=\frac{\kappa-1}{\Rf}.\label{eqn:Sfdinf}
\end{align}
From this we have that 
\begin{align}
    S^\infty=\eta-\Sfd^\infty=\frac{1+a(1-\eta)}{\Rf}
\end{align}
which aligns with the predicted steady state value \cref{eqn:equil2} in \Cref{sec:model} when $\Sfd^*=\Sfd^\infty$. Thus the final fear size is dependent on a new effective reproduction number $\kappa$. Since $\eta$ is the final size of the two susceptible groups during the disease outbreak then $1-\eta$ is the population of those who have recovered. Therefore, fear can spread if the fear reproduction number is sufficiently high such that infecting the remaining susceptibles, $\eta$ with fear is greater than the fear extinction from recovery $a(1-\eta)$. Since $\eta$ and $\Rf$ are independent of $a$ then a fear-endemic state, $\Sfd>0$ occurs when
\begin{align}
    a<a_c;\qquad a_c=\frac{\Rf\eta-1}{1-\eta}.
\end{align}
we demonstrate this in \cref{fig:fearbifa} for parameters $\Rf=8$, $\Ro=2$, $p=0.25$, and $I_0=\delta=0.01$ whence $\eta=0.2197$ and $a_c=0.9708$. If instead we fix $a$ then $\eta=\eta(\Rf)$ and so the bifurcation relationship is non-linear. The bifurcation point, ${\Rf}_c$ for a fear endemic equilibrium in this case from \cref{eqn:Sfdtau} satisfies
\begin{align}
    {\Rf}_c\eta({\Rf}_c)-a(1-\eta({\Rf}_c))=1
\end{align}
and since $\delta\ll 1$ we pose an expansion ${\Rf}_c={\Rf}_{c0}+\delta{\Rf}_{c1}+\dots$ which yields,
    \begin{align}
        {\Rf}_{c0}=\frac{1+a(1-x_0)}{x_0},\quad
        {\Rf}_{c1}=-\frac{(1+a)(1+a(1-x_0))(y_0+x_1)}{x_0^3}.
    \end{align}
For parameters $a=0$, $\Ro=2$, $p=0.25$, and $I_0=\delta=0.01$ we compute that ${\Rf}_c=4.693$ which is demonstrated in \cref{fig:fearbifRf} to good agreement.

\begin{figure}
    \centering
    \subfloat[Infected\label{fig:established_diseasea}]
    {
        \includegraphics[width=0.45\textwidth]{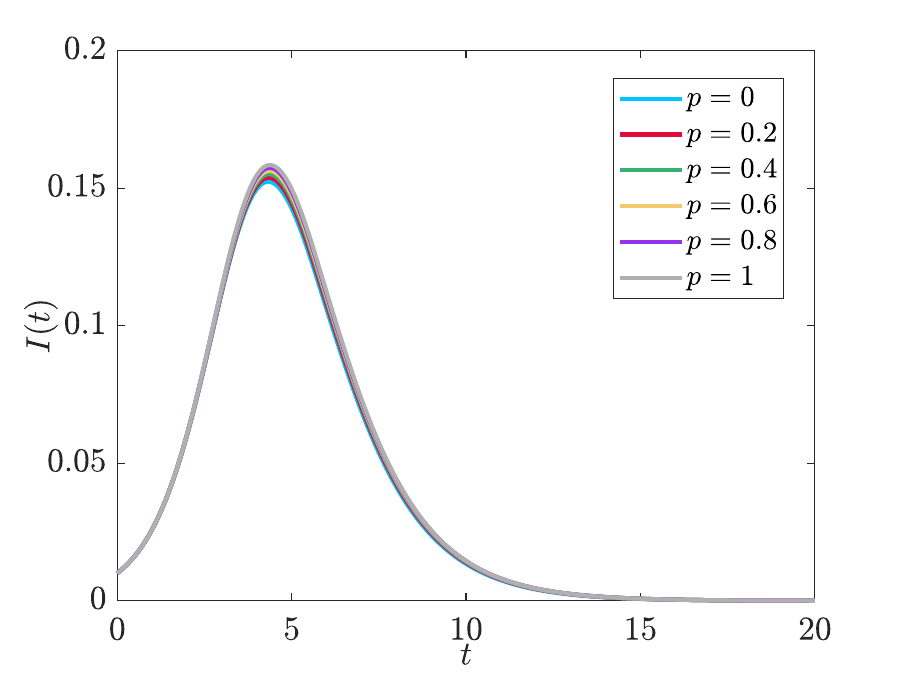}
    }
    \qquad
    \subfloat[Fearful susceptibles\label{fig:established_diseaseb}]
    {
        \includegraphics[width=0.45\textwidth]{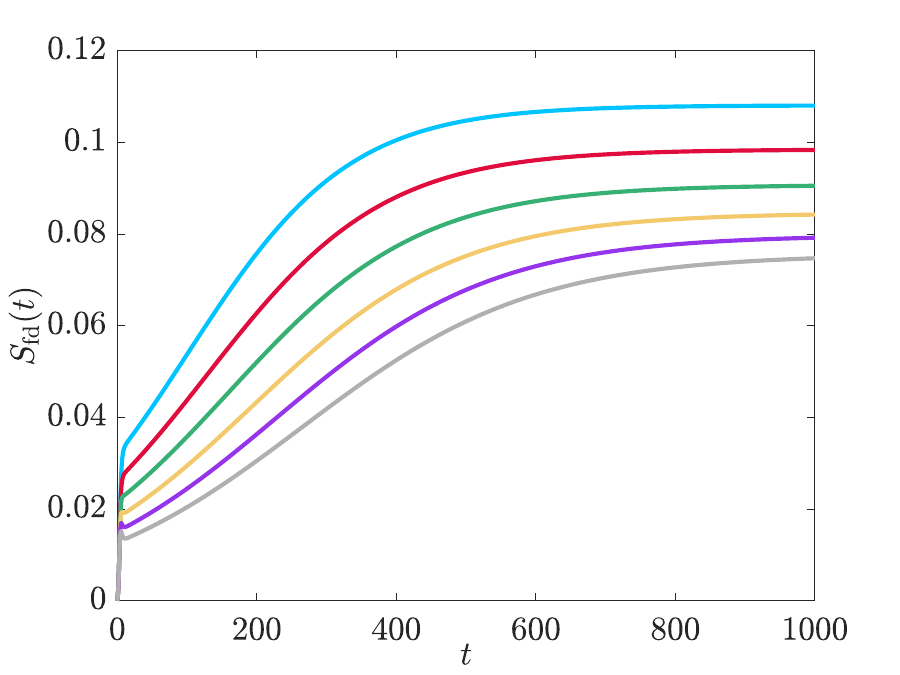}
    }
    \caption{Computation of infected individuals $I(t)$ via simulation of \cref{sys:ndmod} with parameters $\Ro=2$, $\Rf=8$, $a=0$, and $\delta=0.01$ for different values of $p$. From these parameters $\kappa>1$ as defined by \cref{eqn:Sfdtau} and thus the final susceptible group has a fearful proportion. However, $p$ has little impact on the size of the infection. The time scales on the two plots are different indicating that the dynamics of fear settle on a much later timescale than those of the disease.}
    \label{fig:established_disease}
\end{figure}

We emphasize that the role of the long-term limit is to determine the proportions of the susceptible population with and without fear that will remain after the disease is no longer spreading through the population. The reason fear can spread independently of the disease is because susceptible people with fear are able to influence others to become afraid as well. The role of fear is to adopt prophylactic behaviour (reduce contact through $p$), however this parameter has an asymptotically small impact on disease burden (see \cref{fig:established_disease}) after the disease outbreak. Therefore, for diseases with parameters that follow the EDL, the use of fear to alter behaviour has very little impact on disease dynamics and outcomes. However, the long-term limit demonstrates that fear can remain in the system and thus people will adopt prophylactic behaviours, but their efforts will be futile.

\section{The Novel Disease Limit (NDL)}
\label{sec:novel}
We demonstrated in \Cref{sec:established} that the EDL cannot permit any behaviour-related bifurcations in the infection outcomes of disease. We now consider $\Rf=\delta^{-1}\Rfhat$ and $a=\delta^{-1}\hat{a}$ so that we are in the NDL. While numerically observing discontinuity in epidemic prevalence, Epstein et al took parameters $\beta_{\rm fd}\sim\beta$ which in the context of our scaling means $\Rf\gg1$ and thus they were exploring what we now identify as the NDL \cite{epstein2021triple}. For simplicity and without loss of generality we will consider $\Rfhat>\Ro-1$ so that following \cref{eqn:novel_y_growth} the fearful susceptible population grows much faster than the disease. This means that there will be a transient time when the fearless susceptible group rapidly acquires fear of a novel disease before the epidemic grows sufficiently. Therefore, for further simplicity we can consider the initial conditions $S(0)=0$, $\Sfd(0)=1-\delta I_0$, and $I(0)=I_0$ noting that, similar to the EDL in \Cref{sec:established}, a non-zero fearless susceptible population does not impact the qualitative structure of the results. This suggests that we write, $S=\delta x$, $\Sfd=1+\delta y$, and $I=\delta z$ in \cref{sys:ndmod} leading to the model for the NDL
\begin{subequations}\label{sys:novelmod}
    \begin{align}
        \dot{x}=&-\delta\Ro xz-\Rfhat(1+\delta(y+z))x+1+\delta y-\hat{a}(x+y+z)(1+\delta y),\\
        \dot{y}=&-p\Ro(1+\delta y)z+\Rfhat(1+\delta(y+z))x-(1+\delta y)+\hat{a}(x+y+z)(1+\delta y),\\
        \dot{z}=&\left(\Ro p-1+\delta\Ro(x+py)\right)z.\label{eqn:novelmodz}
    \end{align}
\end{subequations}

Once again we pose an expansion
\[
x\sim x_0+\dots,\quad y\sim y_0+\dots, \quad z\sim z_0+\delta z_1+\dots,
\]
where we note that the leading order behaviour in everything except the disease variable will be sufficient. The leading order problem satisfies,
\begin{subequations}\label{sys:novel_leading}
    \begin{align}
        \dot{x}_0=&-\Rfhat x_0+1-\hat{a}(x_0+y_0+z_0),\label{eqn:novelx0de}\\
        \dot{y}_0=&-\Ro p z_0+\Rfhat x-1+\hat{a}(x_0+y_0+z_0),\label{eqn:novely0de}\\
        \dot{z}_0=&(\Ro p-1)z_0,\label{eqn:novelz0de}
    \end{align}
\end{subequations}
with initial conditions $x_0(0)=0$, $y_0(0)=-I_0$, and $z_0(0)=I_0$. The leading order disease dynamics are decoupled from the fear transmission similar to how the fearful susceptible group decoupled from disease dynamics in the EDL of \Cref{sec:established}. This is why we are able to expand $z$ to an additional order since the leading order of $x$ and $y$ will determine the first correction to $z$. Thus,
\begin{align}
    z_0=I_0\exp{(\Rp-1)t};\qquad \Rp=\Ro p.\label{eqn:novelz0}
\end{align}
Adding \cref{eqn:novelx0de,eqn:novely0de} and integrating determines that
\begin{align}
x_0+y_0=-z_0-\hat{z}_0;\qquad \hat{z}_0=\int_0^tz_0(s)\diff s=\frac{I_0}{\Rp-1}\left(\exp{(\Rp-1)t}-1\right),\label{eqn:x0relation}
\end{align}
and this can be used to solve \cref{eqn:novely0de} yielding,
\begin{align}
    y(t)=y_0^\infty-A_0\exp{(\Rp-1)t}-(y_0^\infty-A_0+I_0)\exp{-\Rfhat t},\label{eqn:novely0}
\end{align}
where
\begin{align}
    A_0=\frac{\Rp I_0}{\Rp-1}+\frac{\hat{a}I_0}{(\Rp-1)(\Rp+\Rfhat-1)},\quad y_0^\infty=\frac{(\Rfhat+\hat{a})\frac{I_0}{\Rp-1}-1}{\Rfhat}.\label{eqn:y0infdef}
\end{align}
We have defined $y_0^\infty$ as such since if $\Rp<1$ then $y_0\to y_0^\infty$ as $t\to\infty$. We note that $y_0^\infty<0$ is fine since we have let $\Sfd=1+\delta y$ so the leading order solution is already a correction from unity.

Since the disease is decoupled from fear transmission then the leading order solutions determine the first correction for disease which satisfies,
\begin{align}
    \dot{z}_1=(\Rp-1)z_1-\Ro(z_0+\hat{z}_0+(1-p)y_0)z_0;\qquad z_1(0)=0.
\end{align}
This has solution,
\begin{align}
    z_1(t)=-\Ro\left[U_0\exp{(\Rp-1)t}+U_1\exp{-\Rfhat t}+u^\infty t-(U_0+U_1)\right]z_0,\label{eqn:novelz1}
\end{align}
where
\begin{subequations}
    \begin{align}
        U_0=&\frac{\Rp p I_0}{(\Rp-1)^2}-\frac{\hat{a}(1-p)I_0}{(\Rp-1)^2(\Rp+\Rfhat-1)},\\
        U_1=&\left[\frac{\hat{a}I_0}{\Rfhat^2(\Rp+\Rfhat-1)}-\frac{1}{\Rfhat^2}\right](1-p),\\
        u^\infty=&\frac{\hat{a}(1-p)I_0}{\Rfhat(\Rp-1)}-\frac{pI_0}{\Rp-1}-\frac{1-p}{\Rfhat}.
    \end{align}
\end{subequations}

\begin{figure}
    \centering
    \includegraphics[width=0.6\textwidth]{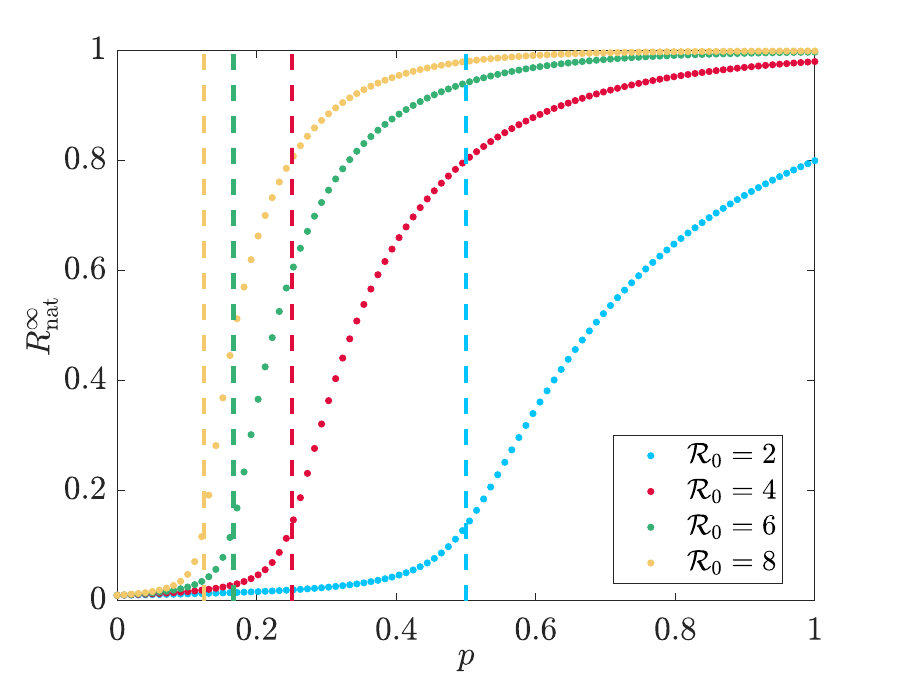}
    \caption{Computation of the final disease burden, $R_{\rm nat}^\infty=\lim_{t\to\infty}R_{\rm nat}(t)$ from simulating \cref{sys:ndmod} with parameters $\hat{a}=0$, $\Rfhat=10$, and $\delta=0.01$ along with initial conditions $S(0)=0$, $\Sfd(0)=0.99$, $I(0)=0.01$. The dashed lines are the predicted bifurcation points $\Rp=1$ from \cref{eqn:novelz0}.}
    \label{fig:Rpbif}
\end{figure}

We remark that in \cref{eqn:novelz0}, yet another effective reproduction number, $\Rp$, emerges. The parameter $p$ is the prophylactic response taken by the fearful susceptible class to reduce disease transmissibility. Thus, if the disease is otherwise causing an outbreak ($\Ro>1$), but the fearful behaviour response causes $\Rp<1$ then the prophylactic behaviour will mitigate disease impact. The leading order response $z_0$ in \cref{eqn:novelz0} is an example of discontinuity in epidemic prevalence as noted by Perra et. al in \cite{perra2011towards} since there will either be exponential growth in the disease leading to a large proportion of disease burden or exponential decay leading to minimal disease burden. Since this is a leading order effect, we do not expect the true discontinuity in final disease burden, but rather a rapid increase in the final recovered population past the predicted bifurcation point $p_c=\Ro^{-1}$. We showcase this bifurcation in Figure \ref{fig:Rpbif} by plotting $R_{\rm nat}^\infty$ from simulating \cref{sys:ndmod} with $\hat{a}=0$, $\Rf=10$, $\delta=0.01$, and varying $p$. We take as initial conditions $S(0)=0$, $\Sfd(0)=0.99$ and $I(0)=0.01$. We choose $\Rfhat$ such that $\Rfhat>\Ro-1$ for all $\Ro$. The recovered group has an inflection point very close to the predicted bifurcation value plotted in dashed lines. 

The equilibrium analysis insures that the only disease steady state for the infection is $I(t)=0$ and therefore unbounded exponential growth cannot happen. The breakdown of the leading order infection \cref{eqn:novelz0} occurs at at time when the exponential becomes large enough such that the assumption $I\sim\mathcal{O}(\delta)$ fails. Therefore, when $\Rp>1$, we would rescale the equations to a long-time limit similar to \Cref{sec:established_long} for the EDL and analyze the final-size in disease burden. However, we instead focus on the second, and more unintuitive, discontinuity in epidemic prevalence associated to the self-limiting behaviour of the complacency parameter $\hat{a}$. This is the unresolved mechanism noted by Perra et. al in \cite{perra2011towards}.

\subsection{Weakly non-linear analysis near $\Ro^{-1}$}\label{sec:weak}

The complacency parameter $\hat{a}$ counters the prophylaxis of fearful susceptibles since recovery from the disease demonstrates a reduction in risk accelerating transfer away from being fearful. Thus, even if $\Rp<1$ it is possible that there exists a value of $\hat{a}$ such that the disease burden does not remain small. We identify the role of the complacency parameter on disrupting the disease burden final size near its natural bifurcation point $p_c=\Ro^{-1}$ by letting 
\begin{align}
    p=\Ro^{-1}-\sqrt{\delta}\hat{p},\label{eqn:phat}
\end{align}
noting that we have taken a minus sign because we are primarily concerned with the case $p<\Ro^{-1}$ ($\hat{p}>0$) since we have already reconciled that when $p>\Ro^{-1}$ the exponential solutions \cref{eqn:novelz0} at leading order blow-up (see \cref{fig:Rpbif}).

The expansion \cref{eqn:phat} is chosen as a distinguished limit of the system as detailed in \cref{app:novel_break}. We also detail in that appendix how taking this expansion requires a long-time analysis on a time scale $t\sim\delta^{-1/2}$ similar to the $\mathcal{O}(\delta)$ timescale needed for the fearful susceptible population in the EDL as discussed in \Cref{sec:established_long}.

To analyze the long-term behaviour of the NDL we rescale time $t=\delta^{-1/2}\tau$. Following \cref{app:novel_break} we must also rescale each of the model variables $S=\sqrt{\delta}X$, $\Sfd=1+\sqrt{\delta}Y$, and $I=\delta Z$ which transforms the model \cref{sys:ndmod} to
\begin{subequations}\label{sys:novel_long}
    \begin{align}
        \sqrt{\delta}X_\tau=&-\Rfhat X-\hat{a}(X+Y)+\sqrt{\delta}(1-\Rfhat XY-\hat{a}(X+Y)Y-\hat{a}Z)+\mathcal{O}(\delta),\label{eqn:novel_longX}\\
        \sqrt{\delta}Y_\tau=&\Rfhat X+\hat{a}(X+Y)-\sqrt{\delta}(1-\Rfhat XY-\hat{a}(X+Y)Y-\hat{a}Z+Z)+\mathcal{O}(\delta),\label{eqn:novel_longY}\\
        Z_\tau=&(\Ro X+Y-\Ro\hat{p})Z+\mathcal{O}(\sqrt{\delta}),\label{eqn:novel_longZ}
    \end{align}
\end{subequations}
with initial conditions $X(0)=Y(0)=0$ and $Z(0)=I_0$.

Since $\delta\ll1$ \cref{eqn:novel_longX,eqn:novel_longY} are both in a quasi-steady limit yielding,
\begin{align}
    Y=-\left(\frac{\Rfhat+\hat{a}}{\hat{a}}\right)X.\label{eqn:Yquasi}
\end{align}
However, this introduces a degeneracy into the system since both equations yield the same result. This is resolved by adding \cref{eqn:novel_longX,eqn:novel_longY},
\begin{align}
    X_\tau+Y_\tau=-Z+\mathcal{O(\sqrt{\delta})},\label{eqn:XYadd}
\end{align}
which after combining with \cref{eqn:Yquasi} and ignoring higher order terms yields,
\begin{align}
    \frac{\Rfhat}{\hat{a}}X_\tau=Z.\label{eqn:XtauZ}
\end{align}
Combining \cref{eqn:Yquasi,eqn:XtauZ,eqn:novel_longZ} allows us to determine an implicit relationship between the fearless susceptible and infectious groups,
\begin{align}
    Z=\frac{\Rfhat}{\hat{a}}\left(-\frac{\Gamma}{2}X^2-\Ro\hat{p}X+\frac{I_0\hat{a}}{\Rfhat}\right);\qquad \Gamma=\frac{\Rfhat-(\Ro-1)\hat{a}}{\hat{a}},\label{eqn:Gammadef}
\end{align}
where the initial conditions $X(0)=0$ and $Z(0)=I_0$ have been used. Further combining \cref{eqn:Gammadef} with \cref{eqn:XtauZ} yields a first order non-linear differential equation for $X$,
\begin{align}
    X_\tau=-\frac{\Gamma}{2}X^2-\Ro\hat{p}X+\frac{I_0\hat{a}}{\Rfhat},\quad X(0)=0.\label{eqn:Xtaude}
\end{align}
with solution,
\begin{align}
    X=-\frac{\Ro\hat{p}}{\Gamma}+\frac{\sqrt{\theta}}{\Rfhat\Gamma}\textrm{tanh}\left(\frac{\sqrt{\theta}}{2\Rfhat}\tau+\hat{C}^+\right);\qquad \hat{C}^+=\textrm{arctanh}\left(\frac{\Ro\Rfhat\hat{p}}{\sqrt{\theta}}\right),\label{eqn:Xsol}
\end{align}
where
\begin{align}
    \theta=\Ro^2\Rfhat^2\hat{p}^2+2I_0\Rfhat\hat{a}\Gamma.\label{eqn:thetadef}
\end{align}

From \cref{eqn:Xsol} we have a classic final-size argument, namely that the final susceptible group remains bounded,
\[
X^\infty=\lim_{\tau\to\infty}X=-\frac{\Ro\hat{p}}{\Gamma}+\frac{\sqrt{\theta}}{\Rfhat\Gamma}\label{eqn:Xinf}
\]
and that the infection extinguishes. However, this is only true so long as $\theta\geq0$. Firstly, from \cref{eqn:thetadef} we note that if $\Gamma>0$ defined by \cref{eqn:Gammadef} then necessarily $\theta>0$ always. Therefore, the solution can only begin to break down when $\Gamma<0$ which occurs if $\hat{a}>\hat{a}_c$ where
\begin{align}
    \hat{a}_c=\frac{\Rfhat}{\Ro-1}.\label{eqn:ac}
\end{align}
However, even if $\hat{a}>\hat{a}_c$ then if $\hat{p}$ is sufficiently large, $\theta$ can remain positive. Thus for $\theta$ to become negative and for the solution \cref{eqn:Xsol} to break down we require $\hat{a}>\hat{a}_c$ and $\hat{p}<\hat{p}_c$ where
\begin{align}
    \hat{p}_c=\frac{1}{\Ro}\sqrt{\frac{2I_0\hat{a}|\Gamma|}{\Rfhat}}.\label{eqn:phatc}
\end{align}

If $\hat{a}>\hat{a}_c$ and $\hat{p}<\hat{p}_c$ then the argument in the hyperbolic tangent of \cref{eqn:Xsol} becomes complex and instead we can write the solution as,
\begin{align}
    X(\tau)=\frac{\Ro\hat{p}}{|\Gamma|}+\frac{\sqrt{|\theta|}}{\Rfhat|\Gamma|}\tan\left(\frac{\sqrt{|\theta|}}{2\Rfhat}\tau-\hat{C}^-\right),\qquad \hat{C}^-=\arctan\left(\frac{\Ro\Rfhat\hat{p}}{\sqrt{|\theta|}}\right).\label{eqn:Xsoltan}
\end{align}
This blows up when $\tau=\tau_c$ given by,
\begin{align}
    \tau_c=\frac{2\Rfhat}{\sqrt{|\theta|}}\left(\frac{\pi}{2}-\hat{C}^-\right),\label{eqn:tauc}
\end{align}
at which point a new rescaling would be needed to resolve the relaxation to equilibrium.

\begin{figure}
    \centering
    \subfloat[$\theta>0$, $p=0.48$, vary $\Rfhat$.\label{fig:apbifa}]
    {
        \includegraphics[width=0.45\textwidth]{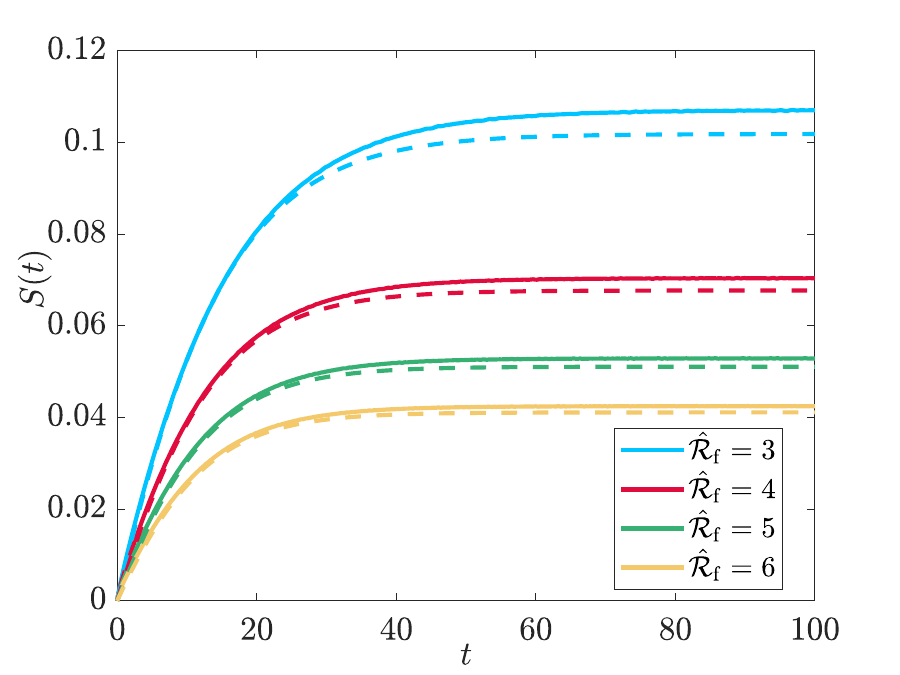}
    }
    \qquad
    \subfloat[$\theta>0$, $p=0.48$, vary $\Rfhat$. \label{fig:apbifb}]
    {
        \includegraphics[width=0.45\textwidth]{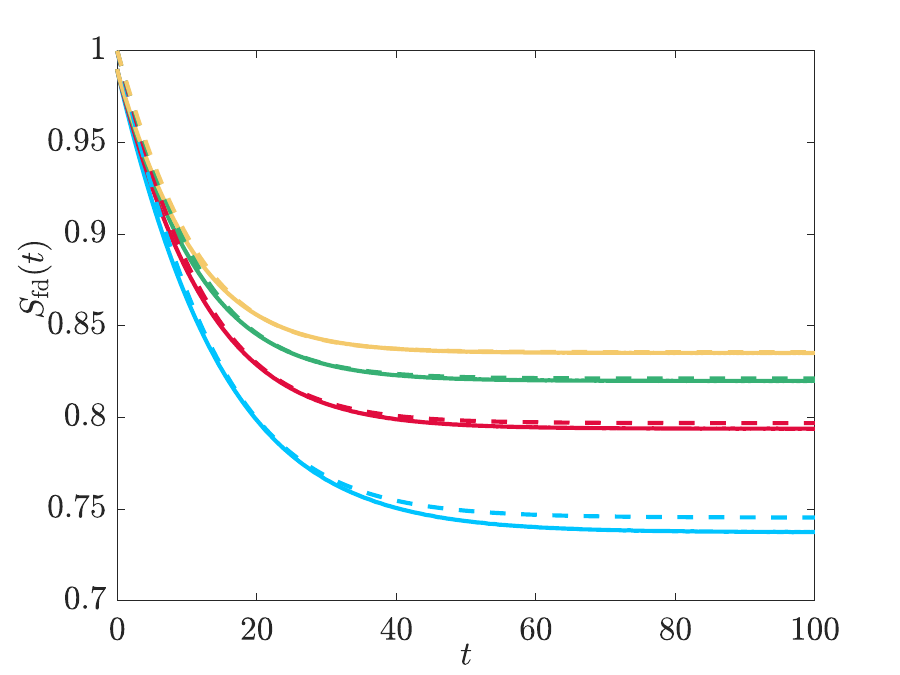}
    }
    \\
    \subfloat[$\theta<0$, $p=0.44$, $\Rfhat=3$. \label{fig:apbifc}]
    {
        \includegraphics[width=0.45\textwidth]{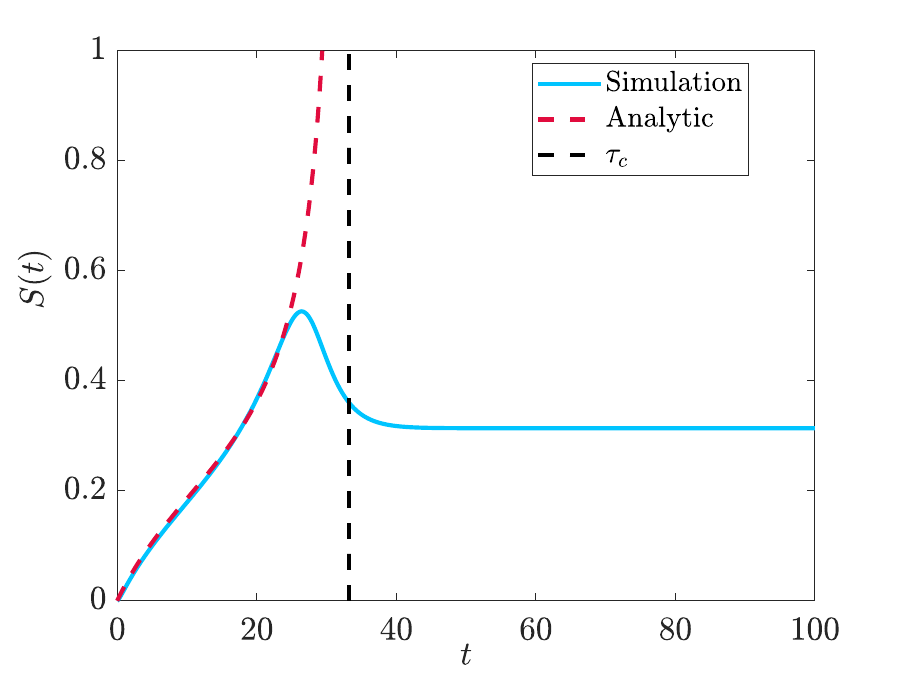}
    }
    \qquad
    \subfloat[$\theta<0$, $p=0.44$, $\Rfhat=3$. \label{fig:apbifd}]
    {
        \includegraphics[width=0.45\textwidth]{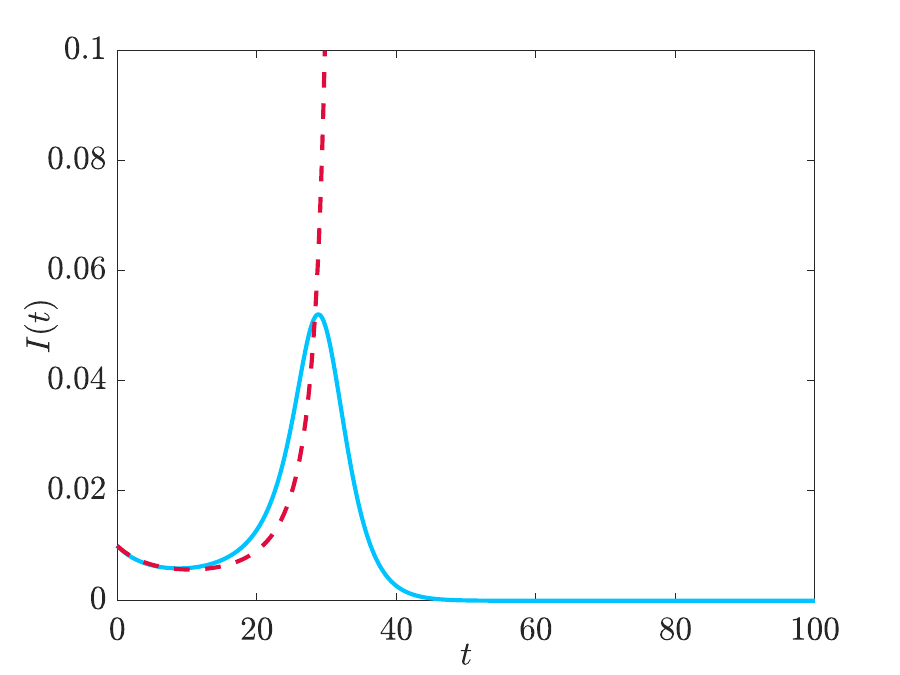}
    }
    \caption{Comparison of simulations of \cref{sys:ndmod} with parameters $\hat{a}=2$, $\Ro=2$, and $\delta=0.01$ along with initial conditions $S(0)=0$, $\Sfd(0)=0.99$, $I(0)=0.01$. The dashed curves are the asymptotic approximation valid on a time scale $t\sim\delta^{-1/2}$ given by \cref{eqn:Xsol} (\cref{fig:apbifa}), \cref{eqn:Yquasi} (\cref{fig:apbifb}), \cref{eqn:Xsoltan} (\cref{fig:apbifc}) and \cref{eqn:Gammadef} (\cref{fig:apbifd}). The vertical black dashed line indicates the approximation $\tau_c$ given by \cref{eqn:tauc} for the time when the asymptotic solution fails if $\theta<0$.}
    \label{fig:apbif}
\end{figure}

We demonstrate the bifurcation behaviour around $\hat{p}_c$ in \cref{fig:apbif} where we simulate \cref{sys:ndmod} with initial conditions $S(0)=0$, $\Sfd(0)=0.99$ and $I(0)=0.01$ while taking $\delta=0.01$. For \cref{fig:apbifa,fig:apbifb} we fix $\Ro=2$, $\hat{a}=2$, and $p=0.48$ ($\hat{p}=0.2$) and vary $\Rfhat$ taking $\Rfhat=3,4,5,6$. This results in $\Gamma=0.5,1,1.5$, and $2$ respectively which are all positive and thus $\theta>0$ defined by \cref{eqn:thetadef}. Therefore, the solutions for the fearless susceptibles should follow $S=\sqrt{\delta}X$ given by \cref{eqn:Xsol} while the fearful susceptibles are given by $\Sfd=1+\sqrt{\delta}Y$ given by \cref{eqn:Yquasi}. We note excellent agreement between the simulation (solid lines) and analytical approximations (dashed lines) improving with increasing $\Rfhat$. This improvement is unsurprising since, for example, when $\Rfhat=3$ then $\Gamma=0.5\sim\sqrt{\delta}$ and so other asymptotic structures are involved.

For \cref{fig:apbifc,fig:apbifd} we fix $\Ro=2$, $\Rfhat=3$, $\hat{a}=8$, and $p=0.44$ ($\hat{p}=0.6$). With these values $\Gamma=-0.625<0$. Furthermore from \cref{eqn:phatc} $\hat{p}_c=0.9129$ and since $\hat{p}<\hat{p}_c$ then $\theta<0$ and so we expect a bifurcation. Therefore, the solutions for the fearless susceptibles now should follow $S=\sqrt{\delta}X$ given now by \cref{eqn:Xsoltan}. Indeed we observe this in \cref{fig:apbifc} where the agreement is excellent until approximately $\tau_c$ (black dashed line) where there is a change in the solution behaviour and the asymptotic approximation fails. We show how this leads to an epidemic in the infected class in \cref{fig:apbifd} where $I=\delta Z$ is given by \cref{eqn:Gammadef}.

\begin{figure}
    \centering
    \includegraphics[width=0.6\textwidth]{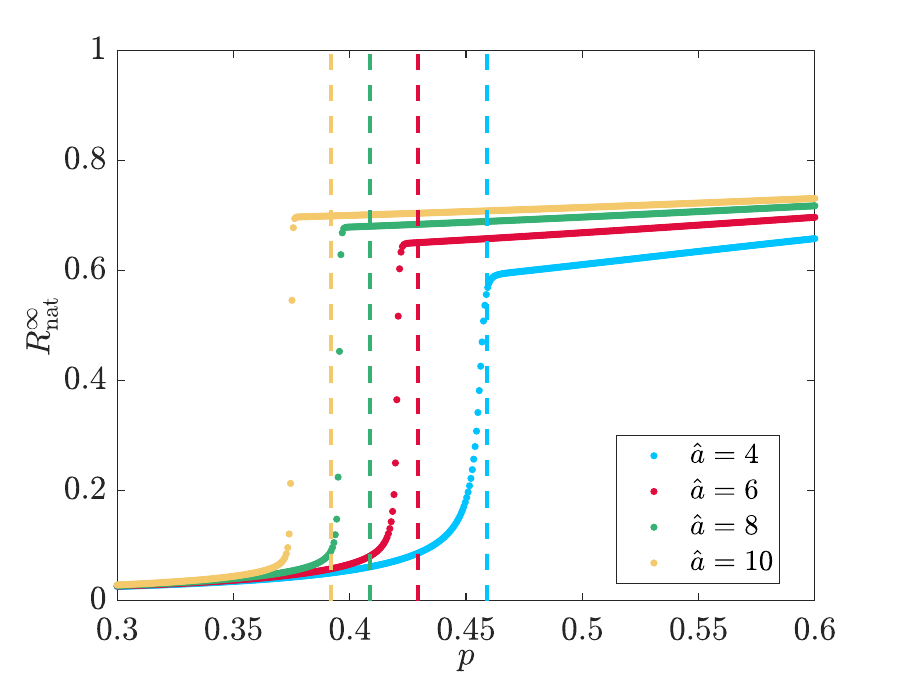}
    \caption{The final epidemic prevalence $R_{\rm nat}^\infty$ simulated from \cref{sys:ndmod} with parameters $\Ro=2$, $\Rfhat=3$ and $\delta=0.01$ along with initial conditions $S(0)=0$, $\Sfd(0)=0.99$, $I(0)=0.01$ while varying $\hat{a}$ and $p$. The dashed curves are the predicted points of discontinuity in epidemic prevalence from \cref{eqn:phatc}.}
    \label{fig:Rdiscont}
\end{figure}

Overall, we have deciphered the mechanism first reported by Perra et al in \cite{perra2011towards} concerning discontinuity in epidemic prevalence. When the prophylactic behaviour of the fearful susceptible population is sufficient to prevent an epidemic ($\Rp<1$) then the self-limiting behaviour of complacency causes an outbreak to emerge when $\hat{a}>\hat{a}_c$ and $p>\Ro^{-1}-\sqrt{\delta}\hat{p}_c$ with $\hat{p}_c$ given by \cref{eqn:phatc} as demonstrated in \cref{fig:apbifd}. We demonstrate the discontinuity in epidemic prevalence in \cref{fig:Rdiscont} simulating \cref{sys:ndmod} with $\Ro=2$, $\Rfhat=3$, and $\delta=0.01$ while varying $\hat{a}$. We plot the predicted bifurcation point $\Ro^{-1}-\sqrt{\delta}\hat{p}_c$ from \cref{eqn:phatc} as vertical dashed lines. We see that the approximations worsen as $\hat{a}$ increases which is expected since, for example if $\hat{a}=10=1/\sqrt{\delta}$ then other asymptotic considerations need to be made.

\begin{figure}
    \centering
    \subfloat[$\delta=0.01$. \label{fig:bifcurvea}]
    {
        \includegraphics[width=0.45\textwidth]{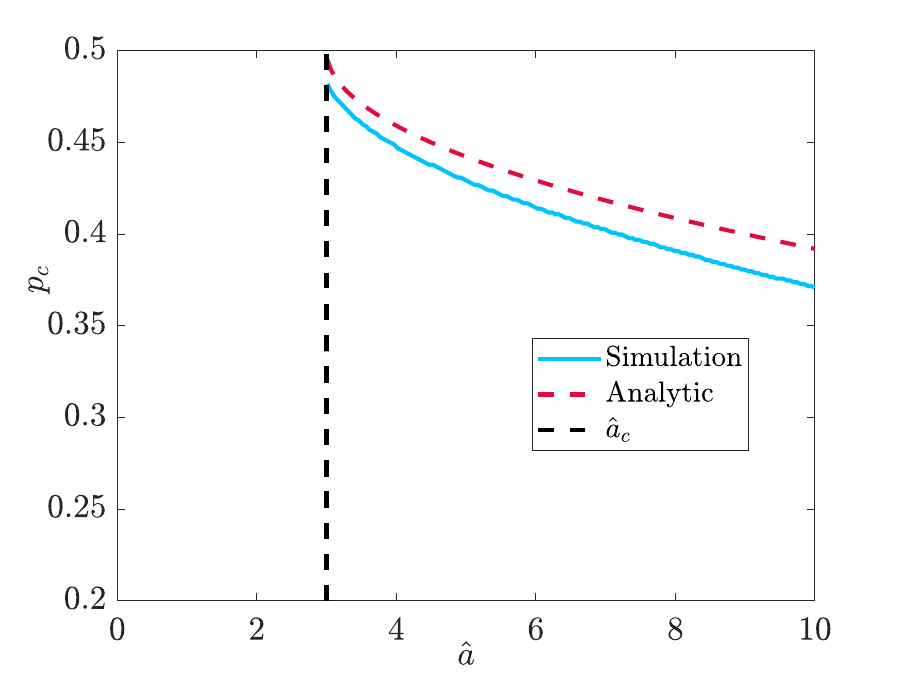}
    }
    \quad
    \subfloat[$\delta=0.001$.\label{fig:bifcurveb}]
    {
        \includegraphics[width=0.45\textwidth]{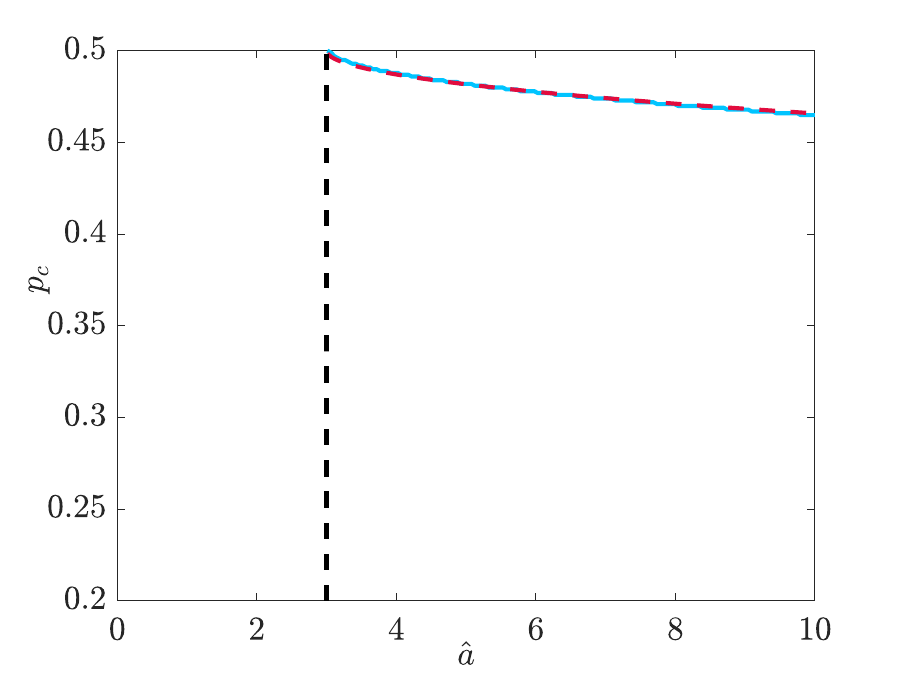}
    }
    \caption{The epidemic prevalence discontinuity bifurcation curve simulated (solid lines) from \cref{sys:ndmod} with parameters $\Ro=2$ and $\Rfhat=3$ along with initial conditions $S(0)=0$, $\Sfd(0)=0.99$, $I(0)=0.01$. The dashed lines are the predicted bifurcation curves given by $p_c(\hat{a})=\Ro^{-1}-\sqrt{\delta}\hat{p}_c$ with $\hat{p}_c$ given by \cref{eqn:phatc}.}
    \label{fig:bifcurve}
\end{figure}

We plot the bifurcation curve $p_c(\hat{a})$ numerically generated from simulating \cref{sys:ndmod} for $\Ro=2$ and $\Rfhat=3$ to that predicted from \cref{eqn:phatc} in \cref{fig:bifcurvea} (for $\delta=0.01$) and  \cref{fig:bifcurveb} (for $\delta=0.001$) showing excellent agreement within $\mathcal{O}(\delta)$, the order considered here.

\begin{figure}
    \centering
    \begin{minipage}[t]{0.45\textwidth}
    	\centering
    	\vspace{0pt}
    	\includegraphics[width=\textwidth]{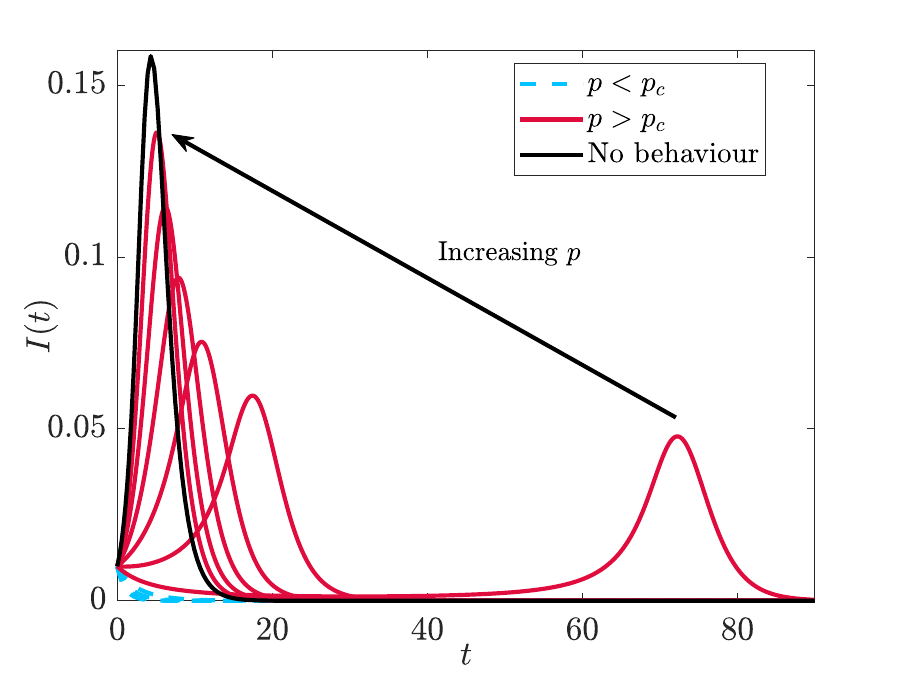}
	    \caption{Simulations of $I(t)$ from \cref{sys:ndmod} with parameters $\Ro=2$, $\Rfhat=3$, $\hat{a}=8$, and $\delta=0.01$ along with initial conditions $S(0)=0$, $\Sfd(0)=0.99$, $I(0)=0.01$. The blue dashed curves are trajectories for $p<p_c$ while the red solid curves are for $p>p_c$ with $p_c=0.4087$ as in \cref{fig:apbif}. The black curve is the simulation of a basic SIR model without behavioural intervention (i.e. the fearful susceptible compartment $\Sfd$ is not present).}
	    \label{fig:diseaseburden}
	\end{minipage}
	\hfill
    \begin{minipage}[t]{0.45\textwidth}
		\centering
		\vspace{0pt}
		\includegraphics[width=\textwidth]{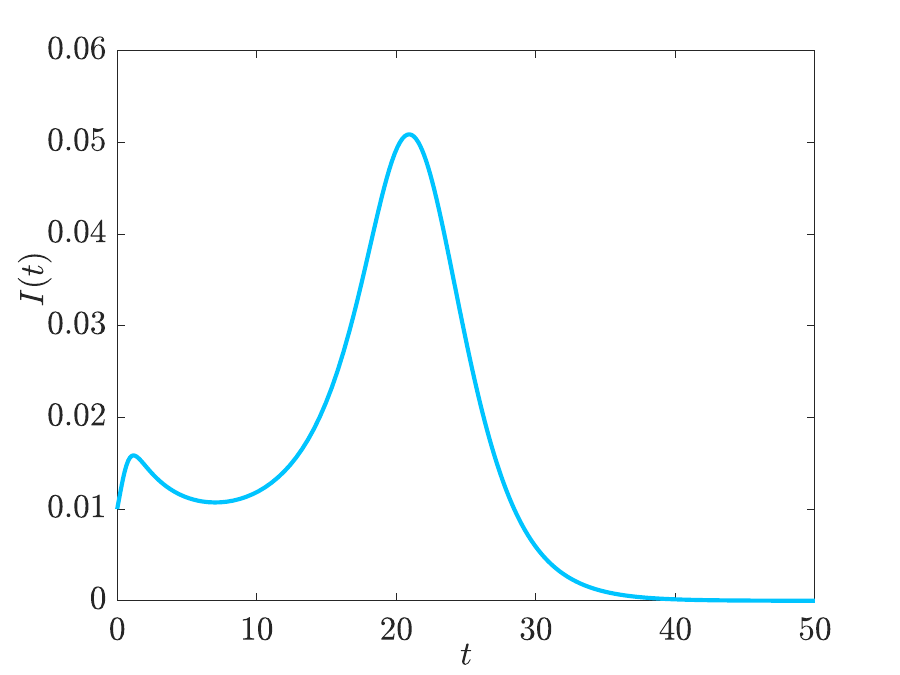}
		\caption{Simulations of $I(t)$ from \cref{sys:ndmod} with parameters $\Ro=2$, $\Rfhat=3$, $\hat{a}=8$, $p=0.41$, and $\delta=0.01$ along with initial conditions $S(0)=0.85$, $\Sfd(0)=0.14$, $I(0)=0.01$. The initial disease peak is because of initial fearless susceptibles and $\Ro>1$ while the secondary peak is the delay from fear complacency as discussed in \cref{sec:weak}.}
		\label{fig:multipeak}
	\end{minipage}	
\end{figure}

Finally, we plot the infected class $I(t)$ varying $p$ in \cref{fig:diseaseburden}. We see that when $p<p_c$ then there is no outbreak as predicted (blue curves). However, when $p>p_c$ a disease outbreak occurs (red curves). We also plot in black the infection curve if behaviour is not included in the model at all (no $\Sfd$ compartment). As $p$ increases past $p_c$ an infection emerges from infinite time moving towards $t=0$ as $p$ increases further coinciding on the behaviour-free black curve when $p=1$. $p=1$ means that the fearful group does not adopt prophylaxis and thus they are indistinguishable from the fearless group. From \cref{fig:diseaseburden} we see that the role of prophylactic behaviour when an outbreak occurs is to delay the outbreak and reduce its severity. The delay is because when adopting prophylaxis, there is limited susceptibility initially. However, as recovery happens there is a transition out of the fearful group which ignites the outbreak. The delay allows for recovery of infection reducing the number of infections when the fully susceptible class emerges leading to the reduced severity. We note that for simplicity, our analysis (and hence simulations) were based on everyone initially being in the fearful susceptible class. If a more proportional mixing was used with some people in the fearless and fearful susceptible classes then, in addition to the delayed infection wave driven by fear, there will be an initial infection wave of the fearless susceptibles, since $\Ro>1$, and thus there will be multiple infection peaks (see \cref{fig:multipeak}) as observed by Epstein et al. \cite{epstein2021triple}. We leave the analysis of multiple waves to future work.

\section{Discussion and Conclusions}
\label{sec:Discussion}
We have considered a behaviour-modified susceptible-infected-removed mathematical model within which disease-susceptible individuals can acquire fear of the disease and therefore reduce their contact behaviour and hence their susceptibility. Such behaviour considerations are important since the traditional SIR model with constant parameters can only permit a single outbreak, however recurrent waves of disease during a single outbreak are often realized \cite{mummert2013}. We observed that the combination of prophylactic and complacent behaviours can lead to multiple disease outbreaks in \cref{fig:multipeak}.

Often the effect of behaviour on disease transmission is captured through estimation of a time-dependent effective reproduction number \cite{gostic2020}. However, estimates of this can vary dependent on the assumptions, data, and statistical methods used to calculated it \cite{brockhaus2023}. Thus, more mechanistic considerations of behaviour can identify predictive ways in which behaviour affects disease transmission through meaningful, identifiable, and measurable parameters as well as a way to inform parameters by incorporating behaviour and other sociological data.

Beginning with a naive and fearless susceptible population introduced to a small initial infection, we showed that two distinct dynamic limits emerge. The first one occurs when the reproduction numbers of fear and disease are comparable meaning that fear is acquired on a timescale much slower than the infection time. We called such a limit the Establish Disease Limit because there is an implication of familiarity with the disease and its consequences since the behaviour response trails increasing case numbers. The second limit was one in which the behaviour response was rapid and preceded growth in case numbers. Correspondingly the fear reproduction number is much larger than that of the disease implying that there is a relatively unknown understanding of the disease burden. As such we called this the Novel Disease Limit.

The explicit inclusion of behaviour in the model allowing for two distinct dynamic limits is very insightful. It shows how important human behaviour is in assessing disease impacts and burden reduction strategy. It also provides an early-warning system for policy makers since data from the onset of an outbreak can be used to estimate model parameters. These parameter values would indicate if the population response was in the Established or Novel disease limits providing insight into mitigation strategies that would be effective.

The EDL analyzed in \Cref{sec:established} showed that a disease outbreak will occur if the disease basic reproduction number $\Ro$ is greater than one. This is an unsurprising result because this limit has a delayed behavioural response. Once the epidemic ends then there is a long-time limit on the timescale of fear recovery $\gamma_{\rm f}^{-1}$ where fear response becomes important. At this point, since the disease spread has ended, the model behaves as a susceptible-infected-susceptible (SIS) model with fear as the ``infection''.

A characteristic result of the SIS model is that the disease either vanishes or becomes endemic~\cite{brauer2019mathematical}. We confirmed this feature in \cref{fig:fearbif} where we showed that the presence of fear at steady state was dependent on the fear reproduction number $\Rf$ and the fear-loss due to complacency, $a$. The endemicity of infection (analogous to fear in our case) implies that a weak fear response can result in people retaining fear even after the outbreak is over. Such an endemic fear state has been observed for example during COVID-19 when people continued wearing masks while the rate of disease spread was slow \cite{kwon2022}.

Since the behavioural response to the disease is delayed in the EDL then the population is not taking prophylactic action when needed at the onset of the outbreak, but are taking it when the disease outbreak is over and it is not needed. We showed in \cref{fig:established_diseasea} that because of this inverted response, prophylactic behaviour has limited ability to mitigate epidemic prevalence in the EDL. Having a fearful susceptible population will reduce any resurgent outbreaks since they will be adopting prophylactic behaviour. However, this may only have value for a short duration between sequential outbreaks as natural and complacency fear removal mechanisms will reduce the fearful population.  There may also be negative consequences with retaining fear in the population. For example, people may observe that their behavioural changes are unimpactful and resist behavioural change in future outbreaks. There may also be civil unrest, particularly if the behavioural change is imposed through government action \cite{cordell2023}.

Through a weakly nonlinear analysis in the NDL presented in \Cref{sec:novel} we were able to resolve the bifurcation observed by Perra et al. in \cite{perra2011towards} that even if the prophylactic strength is such that $\Rp<1$ then it may not be enough to suppress an outbreak. This is possible because, while the role of $p$ is to reduce disease transmission, there is a second parameter, $\hat{a}$ which is a behaviour complacency rate. This causes people to relax their behaviour to normal conditions and hence in the context of the model, lose fear. This rate of fear loss is proportional to the number of people who have recovered, thus as the burden and consequences of the disease become apparent and people begin to get better, there is less urgency to be afraid and people relax. Overall then, there is a competition between mitigating behaviour that reduces outbreaks and complacent behaviour which encourages them. This leads to the discontinuity in epidemic prevalence observed in \cref{fig:Rdiscont}.

Epidemiologically, the discontinuity in epidemic prevalence occurs because when complacency is sufficiently high then a significant proportion of people stop contributing to the prophylactic behaviour. Those that remain must increase the strength of their prophylaxis to overcome this decrease in population. Thus, since $1-p$ represents a contact reduction, $p$ must decrease to continue to prevent an epidemic.

Understanding the tipping point in epidemic prevalence is of great value to public health. If there is an estimate of population prophylaxis and sentiment to adherence behaviour or complacency then our analysis can predict how well the disease is being managed and how close we are to an epidemic outbreak. Understanding the mechanism can also help with the development of public health messaging. If the mitigating behaviour is sufficient such that the reproduction number falls below one then when disease burden continues to climb, people may feel that their efforts have no value and fatigue sets in \cite{petherick2021}. However, our result allows a quantification of the required prophylactic behaviour to suppress disease spread which could encourage adherence.

Having identified two distinct disease limits, it is of interest in future work to consider the transition between novel and established diseases such as when a new disease becomes endemic in the population. In fact this transition in human behaviour could lead to a robust definition of an endemic disease. Finally, we remark that in the model analyzed here, acquisition of fear is directly proportional to the infected class which mimics the mass-action kinetics between susceptible and infected individuals in a standard SIR model. While that assumption can be justified epidemiologically because two individuals in close approximation can spread a disease unaware of the health status of the other, fear requires a cognizant awareness. Thus, the model assumes that if infection is driving fear then either susceptible people know the infection status of those they come in contact with or are being provided accurate real-time information of the number of infections. In reality it is likely that neither of things are true. Therefore, it will be interesting to explore other pathways of fear acquisition including under-reporting of cases or intentional misinformation campaigns.

\appendix

\section{Solution Breakdown in the Established Disease Limit}\label{app:established_break}
For the EDL we solve \cref{sys:etablishedmod} by expanding in an asymptotic expansion for $\delta$,
\[
S\sim x_0+\delta x_1,\quad \Sfd\sim\delta y_0+\delta^2y_1,\quad I\sim z_0+\delta z_1.
\]
The leading order solutions are given by \cref{eqn:SIRsol} and \cref{eqn:y0sol} and as $t\to\infty$ have values $z_0^\infty=0$, $x_0^\infty$ given by \cref{eqn:x0inf} and $y_0^\infty$ given by \cref{eqn:y0sol}. The next order equations are \cref{sys:establishedOdel} which if we consider a long-time where $x_0\to x_0^\infty$, $y_0\to y_0^\infty$, and $z_0\to 0$ then these differential equations are approximately given by
\begin{align}
    \dot{x}_1\approx&-\Ro x_0^\infty z_1,\\
    \dot{y}_1\approx&(\Rf x_0^\infty-\Ro p y_0^\infty)z_1+\Rf x_0^\infty y_0^\infty-y_0^\infty-a(1-x_0^\infty)y_0^\infty,\label{eqn:y1deapprox}\\
    \dot{z}_1\approx& (\Ro x_0^\infty-1)z_1.
\end{align}
The last of these has solution,
\begin{align}
    z_1=C_z\exp{-(1-\Ro x_0^\infty)t},\label{eqn:z1approxsol}
\end{align}
for constant $C_z$ which would be determined from early-time considerations. Since $x_0^\infty<\Ro^{-1}$ then this decays and we can assume in the long-time limit that the disease has gone extinct. We can then solve \cref{eqn:y1deapprox} yielding,
\begin{align}
    y_1=C_y+\frac{(\Ro p y_0^\infty-\Rf x_0^\infty)}{1-\Ro x_0^\infty}C_z\exp{-(1-\Ro x_0^\infty)t}+y_0^\infty\left(\Rf x_0^\infty-1-1(1-x_0^\infty)\right)t, \label{eqn:y1approxsol}
\end{align}
for another constant $C_y$. Problematically, we note that \cref{eqn:y1approxsol} grows without bound due to the linear term. Since the expansion is $y_0+\delta y_1$ then this sequence of terms will no longer be asymptotic when $t\sim\mathcal{O}(\delta^{-1})$. Therefore we need to consider \cref{sys:ndmod} with a new timescale $t=\delta^{-1}\tau$ where $I=0$ to determine the solution behaviour of $S$ and $\Sfd$ in a long-time limit. As $\tau\to0$ these new solutions must match to the solutions for $t\sim\mathcal{O}(1)$ as $t\to\infty$ (cf. \cite{hinch1991}). Therefore,
\begin{align}
\lim_{t\to\infty}S(t)=&x_0^\infty+\delta x_1^\infty\sim\lim_{\tau\to0}S(\tau)\\
\lim_{t\to\infty}\Sfd(t)=&\delta y_0^\infty\sim\lim_{\tau\to0}\Sfd(\tau).
\end{align}
Since these are just constant then this matching is equivalent to taking $S=x_0^\infty+\delta x_1^\infty$ and $\Sfd=\delta y_0^\infty$ when $\tau=0$ which are the initial conditions used in \Cref{sec:established_long}.

\section{Solution Breakdown in the Novel Disease Limit}\label{app:novel_break}

The leading order behaviour in the Novel Disease Limit as discussed in \Cref{sec:novel} has solutions with an exponential argument $(\Ro p-1)$ and thus the solution behaves very differently around a critical point $p=\Ro^{-1}$. To see how this appears more directly in the model, consider starting with a small initial infection $I\sim\mathcal{O}(\delta)$ and looking at a long-time limit to \cref{sys:ndmod}. The first two equations relate the susceptible compartments which produces two possible quasi-steady states,
\begin{align}
    \Sfd=0;\qquad \Sfd=1-\frac{\Rfhat+\hat{a}}{\Rfhat}S\label{eqn:Sfdlong1}
\end{align}
and we will focus on the latter fear-present case since the leading order fearful behaviour in \Cref{sec:novel} given by \cref{eqn:novely0} is non-zero. The quasi-steady equation for infection \cref{eqn:ndmodI} is,
\begin{align}
0=f(S,\Sfd)I;\qquad f(S,\Sfd)=\Ro(S+p\Sfd)-1.\label{eqn:quasiI}
\end{align}

From \cref{eqn:quasiI} the only quasi-steady solution is $I=0$ unless $f(S,\Sfd)=0$ which occurs when
\begin{align}
    \Sfd=\frac{1-\Ro S}{\Ro p}.\label{eqn:Sfdlong2}
\end{align}
This indicates a transition in the dynamics of infection from decay to growth. Combining \cref{eqn:Sfdlong1,eqn:Sfdlong2} then the dynamics of the infection will begin to transition when $S=S^\dagger$ given by
\begin{align}
    S^\dagger=\frac{(\Rp-1)\hat{a}}{\Ro((\Rfhat+\hat{a})p-\hat{a}}.\label{eqn:longtimeS}
\end{align}

Initially in \Cref{sec:novel} we consider, without loss of generality, that $\Sfd\approx1$ and $S\approx0$ and so $f(S,\Sfd)\approx\Ro p-1$. Assuming that $\Ro p<1$ then the long term behaviour of fearful susceptibles is $\Sfd\sim 1+\delta y_0^\infty$ given by \cref{eqn:y0infdef} while the long term behaviour of fearless susceptibles is $S\sim\delta x_0^\infty$ where
\begin{align}
    x_0^\infty = \frac{I_0\hat{a}+1-\Rp}{\Rfhat(1-\Rp)}.
\end{align}
Since the infection is initially decaying then we are interested in the values of $p$ where $S=\delta x_0^\infty=S^\dagger$ given by \cref{eqn:longtimeS} since that is when the infection behaviour can transition to growth. Equating these leads to a rational expression with quadratic numerator,
\begin{align}
    \frac{a_2p^2-a_1p+a_0}{\Ro\Rfhat(\Rp-1)(\Rf p-\hat{a}(1-p))}=0,\label{eqn:pquad}
\end{align}
where
\begin{subequations}
    \begin{align}
        a_0=&\hat{a}(\Rfhat-\delta\Ro(1+I_0\hat{a})),\\
        a_1=&\Ro(2\hat{a}\Rfhat-\delta(\Rfhat+\hat{a}(I_0\Rfhat+\Ro+1)+I_0\hat{a}^2),\\
        a_2=&\Ro^2(\Rfhat\hat{a}-\delta(\Rfhat+\hat{a}).
    \end{align}
\end{subequations}
When $\delta>0$ then \cref{eqn:pquad} has two distinct solutions for $p$. However, when $\delta=0$ then the quadratic has a single degenerate root $p=\Ro^{-1}$. This confirms that $\Rp\approx1$ triggers the bifurcation, however the degenerate nature suggests the corrections are expanded in powers of $\sqrt{\delta}$ \cite{hinch1991}. Thus when analyzing the behaviour near $p=\Ro^{-1}$ we consider this distinguished limit and let,
\begin{align}
    p=\Ro^{-1}-\sqrt{\delta}\hat{p}.\label{eqn:appphat}
\end{align}

We substitute \cref{eqn:appphat} in the leading order behaviour $\Sfd=1+\delta y_0$ and $I=\delta z_0$ with $y_0$ and $z_0$ given by \cref{eqn:novely0,eqn:novelz0} respectively and expand for $\delta\ll1$ yielding,
\begin{subequations}\label{sys:phatexpand}
    \begin{align}
        y_0\sim&-\frac{\Rfhat+\hat{a}}{\Rfhat}I_0t+\frac{I_0\hat{a}}{\Rfhat^2}-\frac{1}{\Rfhat}-I_0-\left(\frac{I_0\hat{a}}{\Rfhat^2}-\frac{1}{\Rfhat}\right)\exp{-\Rfhat t},\\
        z_0\sim&I_0(1-\Ro\sqrt{\delta}\hat{p}t).\label{eqn:phatz0}
    \end{align}
\end{subequations}
Similarly from the relation \cref{eqn:x0relation} then
\begin{align}
    x_0=-y_0-z_0-\hat{z}_0,\qquad \hat{z}_0\underset{\delta\ll1}{\approx}I_0\left(t-\sqrt{\delta}\frac{\Ro\hat{p}}{2}t^2\right).\label{eqn:phatx0relation}
\end{align}
We see in \cref{eqn:phatz0} that the asymptotic sequence breaks down if $t\sim\mathcal{O}(\delta^{-1/2})$ and so we let $t=\delta^{-1/2}\tau$ which we substitute into \cref{sys:phatexpand,eqn:phatx0relation} producing to leading order,
\begin{subequations}
    \begin{align}
        S\sim\delta x_0\approx&\delta^{1/2}\hat{a}I_0\tau,\\
        \Sfd\sim 1+\delta y_0\approx&1-\delta^{1/2}\frac{\Rfhat+\hat{a}}{\Rfhat}I_0\tau,\\
        I\sim\delta z_0\approx&\delta I_0(1-\Ro\hat{p}\tau),
    \end{align}
\end{subequations}
which suggests that for this long-time limit we expand,
\begin{align}
    S\sim\delta^{1/2}X,\quad \Sfd\sim1+\delta^{1/2}Y,\quad I\sim\delta Z.
\end{align}
This is precisely the scaling used in \Cref{sec:weak} to perform the weakly non-linear analysis in the long-time limit near $p=\Ro^{-1}$. Similar to the EDL in \cref{app:established_break}, matching will necessitate that as $\tau\to0$ then $X\sim Y\sim0$ and $Z\sim I_0$.

\bibliographystyle{siamplain}

\end{document}